\documentclass[journal]{IEEEtran}
\usepackage{amsmath,amsfonts,amssymb}
\usepackage{algorithmic}
\usepackage{algorithm}
\usepackage{array}
\usepackage[caption=false,font=normalsize,labelfont=sf,textfont=sf]{subfig}
\usepackage{textcomp}
\usepackage{stfloats}
\usepackage{float}
\usepackage{url}
\usepackage{verbatim}
\usepackage{graphicx}
\usepackage{cite}
\usepackage{color}
\usepackage{hyperref}

\hypersetup{
  colorlinks=true,
  linkcolor=blue,   
  citecolor=blue,   
  urlcolor=blue     
}

\usepackage{mathtools}

\newcommand\Let{\mathrel{\mathop:\!\!=}}
\newcommand\teL{\mathrel{=\!\!\mathop:}}

\begin{document}

\title{Wideband Gaussian Noise Model of Nonlinear Distortions From Semiconductor Optical Amplifiers}

\author{Hartmut Hafermann,~\IEEEmembership{Senior Member,~IEEE}
\thanks{The author is with Optical Communication Technology Lab, Paris Research Center, Huawei Technologies France, 92100 Boulogne-Billancourt, France (e-mail: hartmut.hafermann@huawei.com).
Published in \emph{Journal of Lightwave Technology}. Digital Object Identifier \href{https://doi.org/10.1109/JLT.2025.3643307}{10.1109/JLT.2025.3643307}.}%
}

\IEEEpubid{\begin{tabular}[t]{@{}l@{}}\makebox[\columnwidth]{\copyright~2025 IEEE. All rights reserved, including rights for text and data mining, and training of artificial intelligence and similar technologies.\hfill}\\\makebox[\columnwidth]{Personal use is permitted, but republication/redistribution requires IEEE permission. See \url{https://www.ieee.org/publications/rights/index.html} for more information.\hfill}\end{tabular}}

\maketitle

\begin{abstract}
A wideband Gaussian Noise Model of the nonlinear noise power spectral density is developed for a single semiconductor optical amplifier as described by the Agrawal model. A simple, interpretable closed-form expression is obtained for the nonlinear noise-to-signal ratio of broadband wavelength-division multiplexed signals as a function of the Agrawal model parameters, the amplifier output power and the transmission bandwidth. 
The accuracy of the closed-form expression and its region of validity are assessed in numerical simulations. The error is smaller than 0.1 dB when the product of bandwidth and gain recovery time $\mathbf{B\times\tau_c}$ exceeds 100.
A complete treatment of gain compression is shown to enhance nonlinear noise by a factor $\mathbf{1+P_\text{out}/P_\text{sat}}$ compared to the first-order perturbation theory result.
\end{abstract}


\section{Introduction}
\IEEEPARstart{S}{emiconductor} Optical Amplifiers (SOAs) are a promising amplifier technology for ultra-wideband (UWB) wavelength-division multiplexed (WDM) transmission systems~\cite{Renaudier2017,Zhao2024,Sillekens2025}, or as booster~\cite{Zhao2023} or pre-amplifiers~\cite{Brenot2024,Jamali2025}.
Compared to Erbium-doped fiber amplifiers (EDFAs), they offer large bandwidth in excess of 100 nm and potential advantages in terms of smaller footprint, better energy efficiency, integration and lower cost. Recent advances in SOA design have resulted in noise figures competitive with those of EDFAs~\cite{Brenot2024} and high saturation power in excess of 23 dBm~\cite{Demirtzioglou2023}.

Nonlinear gain compression and response times of a few hundred picoseconds 
cause signal-dependent nonlinear distortions which are detrimental to system performance. Despite significant progress, nonlinear impairments caused by SOAs remain a concern and must be accounted for in SOA and system design and performance prediction.

For signal distortions caused by nonlinear optical fibers, analytical models based on perturbation theory have proven invaluable to assess the impact of nonlinear impairments, calculate system margins, develop simple rules for system design, or for fast quality of transmission (QoT) prediction.
Various modeling efforts over the years have culminated in the well-known Gaussian noise (GN) model for dispersion unmanaged links~\cite{Poggiolini2012,Johannisson2013}. Ref.~\cite{Poggiolini2014} provides a detailed account of the GN model and its applications. Its limitations are now well understood. Its insensitivity to the modulation~\cite{Mecozzi2012,Dar2013,Dar2014} has led to the development of the enhanced GN (EGN) models~\cite{Carena2014,Serena2015,Ghazisaeidi2017}. Ref.~\cite{Ghazisaeidi2019} extended the theory to account for nonlinear distortions due to SOAs as line amplifiers and their interaction with fiber nonlinearities.

While accurate, the implementation of (E)GN models is complex and requires numerical evaluation of multidimensional integrals. 
Closed-form GN~\cite{Semrau2019,Buglia2023,Buglia2024} and EGN models~\cite{Zefreh2020,Zefreh2021,Jiang2024,Poggiolini2025} facilitate fast QoT prediction and speed up UWB system optimization, but are not always easily interpretable.
Simple closed-form results~\cite{Poggiolini2012}, on the other hand, are useful to provide general intuition, preliminary performance estimates or to inform general system design considerations.
Closed-form GN models have, for example, informed link-optimization heuristics, such as the LOGO(N) strategy~\cite{Poggiolini2013}, which assumes incoherent nonlinear noise accumulation across spans, and underpin the widely used ‘3 dB rule’, which states that at the optimum launch power the amplified spontaneous emission (ASE) equals twice the nonlinear noise power~\cite{Poggiolini2014}.
A closed-form model for SOAs has so far been lacking.

\IEEEpubidadjcol

In this paper, we develop a wideband Gaussian noise model for a standalone SOA as described by the Agrawal model. Its physical interpretation is discussed, highlighting similarities and differences compared to the GN model for fiber links. 
We further obtain simple, interpretable closed-form expressions for the nonlinear noise-to-signal ratio (NSR).
They clearly expose the dependence of nonlinear impairments on the Agrawal model and system parameters, including SOA output power, transmission bandwidth and the signal spectral shape. 
Through detailed error analysis for Gaussian signaling, the model is shown to be in excellent agreement with numerical simulations, even deep in the nonlinear regime, while first-order perturbation theory would underestimate nonlinear NSR by a factor as large as 3 dB at saturation. To fully account for the effect of nonlinear gain compression, it is necessary to sum the deterministic terms in the perturbation expansion of the GN model up to infinite order. In the classic GN model, the corresponding terms only cause an overall phase rotation that does not affect the nonlinear noise.

The restriction to the idealized Agrawal model does not impose a major limitation. While the assumption of a linear dependence of the material gain on carrier density underlying the classical Agrawal model is not suitable for modern multi-quantum well SOAs~\cite{DeTemple1993,Sobhanan2022}, it has been shown that the model still provides a quantitative description of the nonlinear impairments of real devices~\cite{Hafermann2024}, given that its parameters are understood as effective parameters that depend on the respective amplifier working point~\cite{Hafermann2025a}.
It turns out that this even remains true when the wavelength-dependence of gain compression is taken into account. The theory developed here thus remains applicable. The 3 dB-rule, however, no longer holds, and the optimal launch power must in general be computed numerically. Modulation can be accounted for through a modulation-dependent factor~\cite{Hafermann2025b}.

When trends of the Agrawal parameters can be inferred from measurements on multiple SOA designs~\cite{Brenot2024} or otherwise be modeled, the relative performance of different SOA designs in a wideband transmission system can be linked to SOA design parameters through the closed-form GN model. This can inform the SOA design optimization process and allow SOA designers to obtain better tradeoffs.

The remainder of the paper is organized as follows. Section~\ref{sec:agrawal} briefly discusses the Agrawal model. Section~\ref{sec:gn} gives the GN model formula for the noise power spectral density (PSD) in integral form. Section~\ref{sec:closedform} introduces the closed-form GN-model. It is compared to numerical simulations of the Agrawal model in Section~\ref{sec:wdmresults} and its accuracy is assessed in Section~\ref{sec:error}. 
Section~\ref{sec:discussion} discusses the model in light of coherent noise accumulation in transmission links. Section~\ref{sec:conclusions} concludes the paper.

\section{Agrawal's model}
\label{sec:agrawal}

To set the stage for the following discussions, we briefly review the main features of the Agrawal model~\cite{Agrawal1989}.
The propagation of the electric field $E(z,t)$ in an SOA waveguide can be described by
\begin{align}
\frac{\partial E(z,t)}{\partial z} =& \frac{1}{2}\left(g(z,t)-\alpha\right)E(z,t) - j\frac{1}{2}\alpha_H g(z,t) E(z,t),
\label{eq:agrawaldedz}
\end{align}
where $\alpha_H$ is the linewidth enhancement or Henry factor~\cite{Henry1982}, $\alpha$ describes the internal losses of the mode and $g(z,t)$ is the material gain coefficient. The latter implicitly depends on distance and time through the local time-dependent carrier density, $g(z,t)=g(N(z,t))$.

\begin{figure}[!b]
\centering
\includegraphics[width=.95\columnwidth]{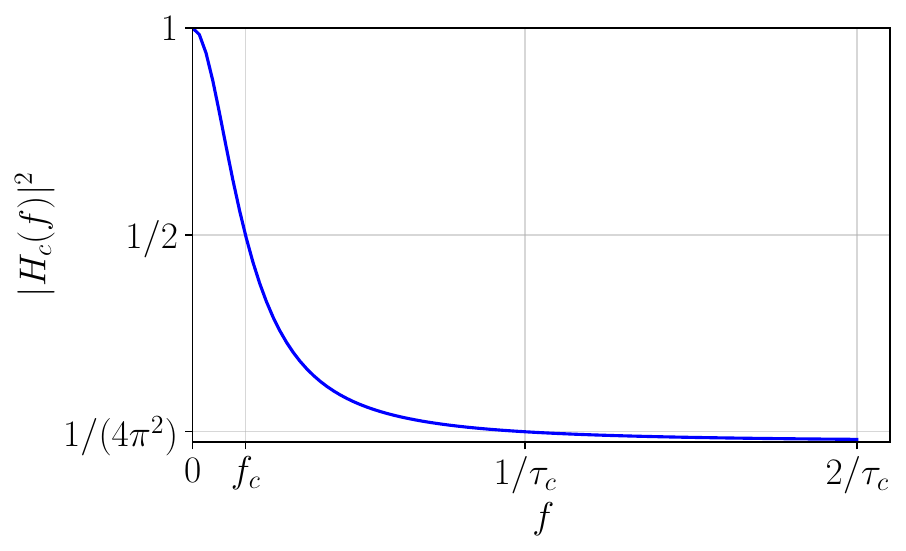}
\caption{Absolute square of the low-pass filter $H_c(f)=1/(1+if/f_c)$. The cutoff frequency is defined as $f_c=1/(2\pi\tau_c)$.}
\label{fig:hc}
\end{figure}

Under the assumption that the carrier recombination rate and the material gain coefficient are both \emph{linear} functions of the carrier density, the rate equation for the carrier density can be recast into the following equation governing the dynamics of the material gain coefficient~\cite{Agrawal1989}:
\begin{align}
\frac{\partial g(z,t)}{\partial t} 
&=\frac{g_0-g(z,t)}{\tau_c} - \frac{1}{\tau_c}\frac{g(z,t)P_\text{in}(t)}{P_\text{sat}}.
\label{eq:gainagrawal}
\end{align}
Here $g_0$ and $P_\text{sat}$ are the small-signal gain coefficient and the saturation power, respectively, which govern the strength of the nonlinear gain compression of the amplifier, and $\tau_c$ is its carrier lifetime. It is commonly assumed that internal losses can be neglected compared to the gain, in which case~\eqref{eq:agrawaldedz} implies $\partial P/\partial z=gP$. Combining with~\eqref{eq:gainagrawal}, the dependence on the spatial coordinate can be integrated out. Specifically, defining $h_0=g_0 L$ and the integrated gain coefficient $h(t)$,
\begin{equation}
h(t) \Let \int_0^L g(z,t) dz,
\label{eq:hdef}
\end{equation}
the equation describing the nonlinear gain dynamics becomes
\begin{align}
\frac{d h(t)}{d t} = \frac{h_0-h(t)}{\tau_c} - \frac{1}{\tau_c}\frac{P_\text{in}(t)}{P_\text{sat}}[\exp(h(t))-1].
\label{eq:hagrawal}
\end{align}
Here $G(t)\equiv\exp\left(h(t)\right)$ equals the time-dependent gain of the amplifier, such that $P_\text{out}(t) = P_\text{in}(t)G(t)$.

It is instructive to consider the effect of the finite carrier lifetime.
For illustrative purposes, consider the case of large gain $G\gg 1$, or $P_\text{in}(t)\ll P_\text{out}(t)$. These conditions are often satisfied in 
applications for optical amplification which require gains in excess of 20 dB, but we emphasize that we do not make such assumption elsewhere in the paper.
Rearranging~\eqref{eq:hagrawal}, we then obtain
\begin{align}
\left(1+\tau_c\frac{d}{d t}\right)(h(t) - h_0) =  - \frac{P_\text{out}(t)}{P_\text{sat}}.
\label{eq:hagrawal2}
\end{align}
In the static case $(dh/dt=0)$, it follows that the gain compression is described by
\begin{align}
G(P_\text{out}) = G_0 \exp\left(-\frac{P_\text{out}}{P_\text{sat}}\right),
\label{eq:gaincurve}
\end{align}
where $G_0\Let\exp(h_0)$ is the small-signal gain.
On the other hand, taking the Fourier transform of~\eqref{eq:hagrawal2}, we find that
\begin{align}
\Delta h(f) =  -H_c(f) \frac{P_\text{out}(f)}{P_\text{sat}},
\label{eq:hagrawal3}
\end{align}
where we have defined $\Delta h(t)\Let h(t)-h_0$ and where
$H_c(f) = 1/(1+i f/ f_c)$ is a low-pass filter determined by the carrier lifetime $\tau_c$ with cutoff frequency
\begin{align}
f_c\Let \frac{1}{2\pi\tau_c}.
\label{eq:fc}
\end{align}
The dynamic gain fluctuations $\Delta h(t)$ are hence proportional to the low-pass filtered output power. The absolute square of the filter function, $|H_c(f)|^2$, is depicted in Fig.~\ref{fig:hc}. It decays from its maximum value of 1 at $f=0$ to 0 in the limit $f\to\infty$. It attains a value of $1/2$ at the cutoff frequency.

It follows from~\eqref{eq:agrawaldedz} that once the gain dynamics $h(t)$ are obtained from~\eqref{eq:hagrawal} through numerical integration, the electric field at the SOA output is computed as
\begin{align}
E_\text{out}(t) = E_\text{in}(t)\exp\left(\frac{1}{2}(1-j\alpha_H)h(t)\right).
\label{eq:field}
\end{align}
Equations~\eqref{eq:hagrawal} and~\eqref{eq:field} are commonly referred to as the Agrawal model. 
As the gain $h(t)$ fluctuates around its average value, the signal experiences amplitude fluctuations as well as phase fluctuations. 
While an optical signal in a fiber primarily experiences phase fluctuations due to the nonlinear Kerr effect, with amplitude fluctuations emerging via phase–amplitude coupling through dispersion, an SOA induces amplitude fluctuations via gain compression that are converted to phase fluctuations through the Henry factor $\alpha_H$~\cite{Henry1982}. In the GN model these impairments are treated as additive Gaussian noise.

\section{GN Model}
\label{sec:gn}

Under the assumption of Gaussian signal statistics and based on Eqs.~\eqref{eq:agrawaldedz} and~\eqref{eq:gainagrawal}, the following Gaussian Noise model expression for the power spectral density of the nonlinear interference noise of an SOA can be obtained:
\begin{align}
G_\text{NLI}(f) =& 
\frac{1}{4}\left(1+\alpha_H^2\right) \frac{P_\text{out}}{1+\frac{P_\text{out}}{P_\text{sat}}}\left(\frac{P_\text{out}}{P_\text{sat}}\right)^2\left(1-\frac{1}{G}\right)^2\notag\\
&\times\!\int_{-\infty}^{+\infty}\!\!\! df_1 \int_{-\infty}^{+\infty}\!\!\! df_2 g_\text{in}(f_1)g_\text{in}(f_2)g_\text{in}(f_1+f_2-f)\notag\\
&\times\!\left[|H_c(f-f_2)|^2\! +\! H_c(f-f_2) H_c^*(f-f_1)\right].
\label{eq:gnlifinal}
\end{align}
where $g_\text{in}(f)\Let G_\text{in}(f)/P_\text{in}$ is the normalized input power spectral density. The derivation is provided in Appendix~\ref{sec:app:gn}. $G$ denotes the static compressed gain at the average total output power $P_\text{out}$ and is determined through the implicit relation:
\begin{equation}
G = G_0\exp\left(-\Big(1-\frac{1}{G}\Big) \frac{P_\text{out}}{P_\text{sat}}\right).
\label{eq:gaincompression}
\end{equation}
The exact solution $G(P_\text{out})$ is provided in Appendix~\ref{app:gaincompression}.

\subsection{Physical interpretation}

\begin{figure}[!t]
\centering
\includegraphics[width=\columnwidth]{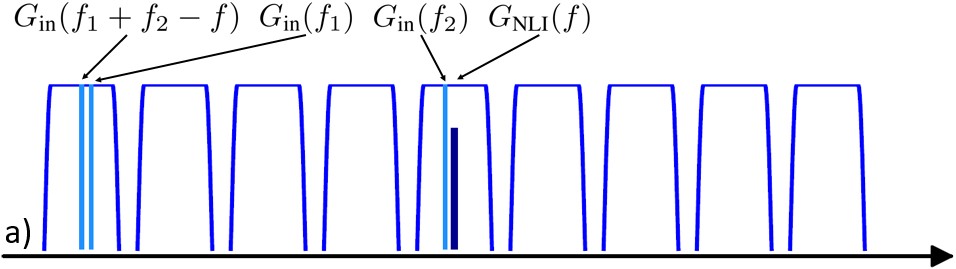}\\
\vspace{1em}
\includegraphics[width=\columnwidth]{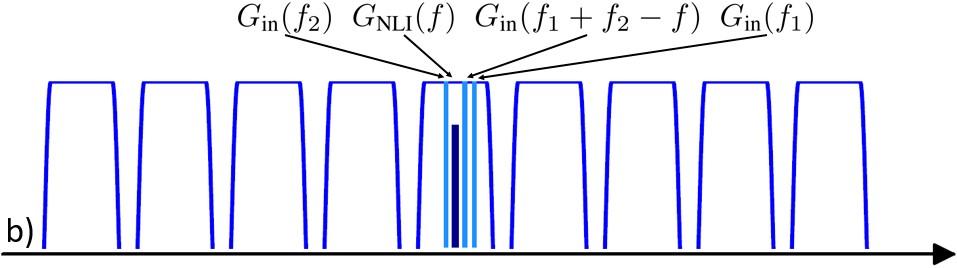}
\caption{Schematic representation of the input power spectral density $G_\text{in}(f)$ and the spectral origin of the dominant contributions to the nonlinear interference noise power spectral density $G_\text{NLI}(f)$. a) Cross-channel interference (XCI). b) Self-channel interference (SCI).}
\label{fig:spectrum}
\end{figure}

The interpretation of the GN model formula~\eqref{eq:gnlifinal} for the Agrawal model bears some similarity to the classic GN model, but also important differences. First consider the spectral origin of the nonlinear noise as described by the second and third line of~\eqref{eq:gnlifinal}. Just like the Gaussian Noise Reference Formula~\cite{Poggiolini2014},~\eqref{eq:gnlifinal} contains a double integral over a product of input power spectral densities at three different frequencies. It can be interpreted as an integral over infinitesimal FWM contributions, as illustrated in Fig.~\ref{fig:spectrum}. 
Each such contribution is weighted by certain products of two filter functions $H_c(f)$ in the third line of~\eqref{eq:gnlifinal}. 

The first term of those is the dominant term as will be shown in Section~\ref{sec:closedform}.
Physically, this term describes the beating of two arbitrary spectral tones at $f_1$ and $f_1+f_2-f=f_1-\Delta f$ that induces an oscillation of the carrier density at the beating frequency $\Delta f = f-f_2$. This oscillation in turn modulates the carrier density-dependent material gain coefficient at this frequency. The nonlinear noise originates from the interaction of a third spectral tone at $f_2$ with the modulated gain, which creates a new tone at $f=f_2+\Delta f$ according to the FWM process. In the GN model, these FWM contributions are integrated and treated as a source of noise.

Because of the finite SOA response time, only beating frequencies of the order of the cutoff frequency or below contribute significantly to the SOA response as described by the filter term $|H_c(f-f_2)|^2=|H_c(\Delta f)|^2$. This term decays with frequency according to Fig.~\ref{fig:hc}. Therefore it only contributes significantly if the beating frequency $\Delta f=f-f_2$ is not much larger than the cutoff frequency $f_c$. 
For typical values of $\tau_c=100\, (200)$ ps, $f_c\approx 1.6\, (0.8)$ GHz, which is much smaller than the typical channel spacing of current transmission systems of $B_\text{ch}=75$ GHz or more. 
Therefore two spectral lines causing significant gain oscillations usually lie within the same channel (or at adjacent edges of neighboring channels). Since the filters in this term do not depend on $f-f_1$, pairs of such beating tones at $f_1$ and $f_1-\Delta f$ in any part of the spectrum contribute equally and are accumulated in the integral over $f_1$ in~\eqref{eq:gnlifinal}.

Following the terminology of~\cite{Poggiolini2012}, the nonlinear noise contributions may be categorized as either cross-channel interference (XCI), when the spectral components $f_1$ and $f_1-\Delta f$ lie in a channel distinct from the channel of interest (COI) containing the frequencies $f$ and $f_2$ as illustrated in Fig.~\ref{fig:spectrum} a), or as self-channel interference (SCI), when the three interacting spectral components lie in the same channel as $f$, as shown in Fig.~\ref{fig:spectrum} b).
In the case of XCI, only the first term in the third line of~\eqref{eq:gnlifinal}, $|H_c(f-f_2)|^2$, contributes significantly, since the second term decays with frequency separation $f-f_1$, which is typically much larger than $f_c$ (unless $f$ and $f_1$ are located at neighboring channel edges). 
Note that since the first term does not depend on $f_1$, XCI does not decay with increasing channel separation, contrary to the case of fiber. In the case of SCI, both terms in the third line of~\eqref{eq:gnlifinal} may contribute substantially.

All three spectral lines may also lie in distinct channels, causing multi-channel interference (MCI). In this case however, both $f-f_2$ and $f-f_1$ are typically much larger than $f_c$, so that both $H_c(f-f_2)$ and $H_c^*(f-f_1)$ are small. 
Therefore MCI is typically negligible for SOAs for typical channel spacings, as in fiber~\cite{Poggiolini2014}. The underlying mechanisms, however, are different. The GN model for fiber contains phase factors of the form $\exp(j4\pi^2(f_1-f)(f_2-f)\beta_2 z)$~\cite{Poggiolini2014}, where $\beta_2$ is the group velocity dispersion parameter and $z$ the distance along the fiber. In regions of large phase mismatch, where both $f_1$ and $f_2$ are far from $f$, the rapid oscillations of the phase factor in the integral tend to cancel out any interference. Only those frequencies for which the phase mismatch is small contribute appreciably to the integral.
There is no analogue of the phase matching condition in an SOA. It is instead the low-pass filtering property due to the finite response time which quenches MCI.

Next, consider the prefactor of the GN model formula in the first line of~\eqref{eq:gnlifinal}.
The contribution proportional to $\alpha_H^2$ is due to phase noise, since gain fluctuations translate to phase fluctuations via the Henry factor $\alpha_H$ according to~\eqref{eq:field}. Due to amplitude fluctuations, the noise power is nonzero even in absence of phase fluctuations ($\alpha_H=0$).
Typical values for $\alpha_H$ found in the literature~\cite{Henry1982,Sobhanan2022,Hafermann2024} range from $3$ to $20$, so that $\alpha_H^2\gg 1$ and phase noise typically dominates significantly over amplitude noise.
The gain factor $1-1/G$ accounts for the fact that gain compression and nonlinear response are driven by the total rate of photons generated by the amplifier, which is proportional to $P_\text{out}-P_\text{in} = P_\text{out}(1-1/G)$, rather than to the output power itself. For line amplifiers with a typical gain of the order of 20 dB, $1/G\ll 1$ and this factor is close to 1.

Finally considering the power dependence of the nonlinear noise power spectral density, one can note that for an amplifier, the nonlinear noise PSD is naturally a function of output power, while for a fiber it is a function of input power.
For a fiber link, nonlinear noise power $P_\text{NL}$ is proportional to $P_\text{in}^3$. The nonlinear NSR can thus be written as $\text{NSR}_\text{NL}=\eta P_\text{in}^2$, where $\eta$ is independent of power~\cite{Poggiolini2014}. On the contrary, for an SOA described by the Agrawal model, the analogous expression as a function of the total SOA output power reads
\begin{equation}
\text{NSR}_\text{NL} = \eta'\, \frac{1}{1+\frac{P_\text{out}}{P_\text{sat}}}\, P_\text{out}^2,
\label{eq:nsr}
\end{equation}
where $\eta'$ has been defined to also be independent of power. Only in the limit of small output power, $P_\text{out}/P_\text{sat}\ll 1$, $\text{NSR}_\text{NL}$ scales as $P_\text{out}^2$. Although much less relevant in practice, for $P_\text{out}/P_\text{sat}\gg 1$, $\text{NSR}_\text{NL}$ becomes proportional to $P_\text{out}$.
Note that because of gain compression described by~\eqref{eq:gaincompression}, $P_\text{out} = G P_\text{in}$ increases more slowly than $P_\text{in}$ when approaching saturation. This will be discussed in more detail in Section~\ref{sec:wdmresults}.

The factor $1/(1+P_\text{out}/P_\text{sat})$ is a direct consequence of nonlinear gain compression. 
As shown in Appendix~\ref{sec:app:heuristic}, a perturbative treatment up to \emph{first order} instead results in a factor $1/(1+P_\text{out}/P_\text{sat})^2$~\cite{Ghazisaeidi2019}, which hence \emph{underestimates} nonlinear noise power by a factor $1+P_\text{out}/P_\text{sat}$. While this factor varies slowly with power relative to $(P_\text{out}/P_\text{sat})^2$, $\text{NSR}_\text{NL}$ is larger by a factor of 2 at saturation compared to the result of first-order perturbation theory. Fully accounting for nonlinear gain compression leads to an \emph{increase} in nonlinear noise power.
It is shown in Appendix~\ref{sec:app:gn} (see in particular the discussion around Eq.~\eqref{eq:infiniteorder}) that nonlinear gain compression arises from the deterministic terms in the perturbation series that are present at \emph{all} orders. The correct power dependence~\eqref{eq:nsr} is obtained by summing the corresponding perturbation series to infinite order.
This term has no analogue in fiber. In a fiber, the corresponding infinite series of deterministic terms leads to an overall nonlinear phase rotation. It is usually not of interest because it neither affects the nonlinear noise, nor is it relevant for transmission because of carrier phase estimation.

Because of the form~\eqref{eq:nsr} of the nonlinear $\text{NSR}$, the so-called '3 dB-rule', which states that at optimum launch power, the 
ratio of amplified spontaneous emission (ASE) to nonlinear noise power equals two, no longer applies.
In addition, it has been shown that for modern multi-quantum well SOAs, the conventional Agrawal model should be replaced by an \emph{effective} Agrawal model, whose parameters depend on the amplifier working point, i.e., on $P_\text{out}$~\cite{Hafermann2025a}. The optimum launch power must thus be found numerically.

\section{Closed-form GN model for WDM signals}
\label{sec:closedform}

\begin{figure}[!t]
\centering
\includegraphics[width=\columnwidth]{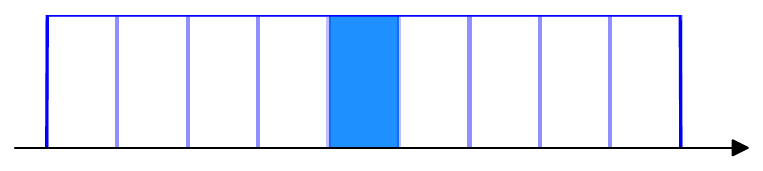}
\caption{Schematic representation of the idealized WDM signal with flat input power spectral density $G_\text{in}(f)$ used for the calculation of the nonlinear interference noise, with the channel of interest indicated by the shaded region.}
\label{fig:spectrumideal}
\end{figure}

\begin{figure}[!t]
\centering
\includegraphics[width=0.818181\columnwidth]{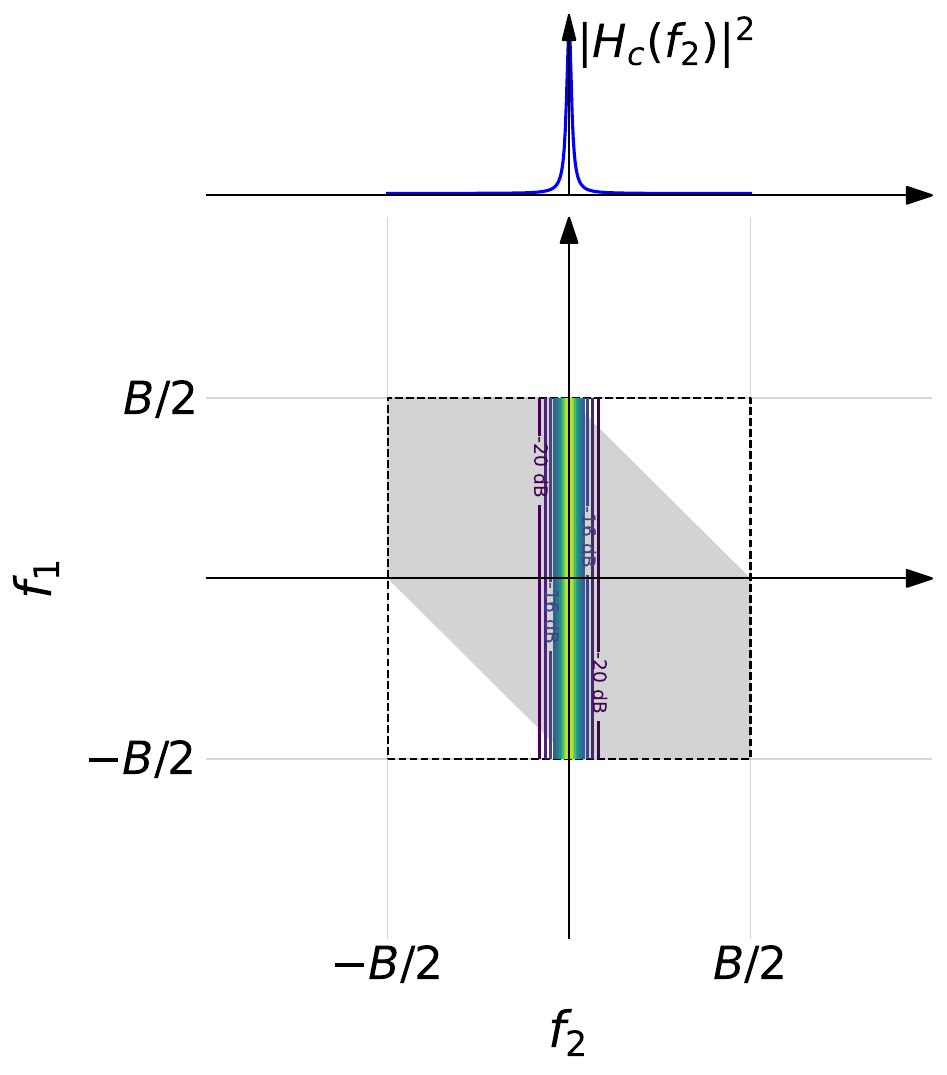}
\caption{Schematic representation of the integration region where the product $g_\text{in}(f_1)g_\text{in}(f_2)g_\text{in}(f_1+f_2)$ is non-zero (gray shaded area) and the extended integration region (dashed square) used in computation of the two-dimensional integral~\eqref{eq:gnlifinal} for the first contribution to the closed-form approximation proportional to $|H_c(f_2)|^2$ ($\tau_c=100$ ps and $B\tau_c=20$). Contour lines of the relative magnitude of this function are overlaid.
Because of the rapid decay of this function, the extension of the integral domain essentially does not change the value of integral. For larger product $B\tau_c$ the decay is even faster.
}
\label{fig:integrala}
\end{figure}

In order to arrive at a closed-form result for the nonlinear noise PSD, we consider the idealized case of a Gaussian signal with a rectangular spectrum of bandwidth $B$, as illustrated in Fig.~\ref{fig:spectrumideal}. Each channel has a frequency spacing $B_\text{ch}$ that coincides with its symbol rate $R_s$ (referred to as ideal Nyquist-WDM in Ref.~\cite{Poggiolini2014}). We assume the center frequency $f_0$ of the COI not to be too close to the band edges (i.e., with the distance well exceeding $f_c$) to avoid edge effects. This assumption is not very restrictive since, as mentioned previously, $f_c$ is rather small. Without loss of generality, one may choose $f_0=0$ to be in the center of the band.
The normalized power spectral density $g_\text{in}(f) = G_\text{in}(f)/P_\text{in}$ then equals one for $f\in[-B/2,B/2]$ and zero otherwise. We first consider the term in the integrand proportional to $|H_c(f_2)|^2$ in~\eqref{eq:gnlifinal}. The integral domain for the two-dimensional integral is the elongated hexagon shown in Fig.~\ref{fig:integrala} (gray shaded area), where the product $g_\text{in}(f_1)g_\text{in}(f_2)g_\text{in}(f_1+f_2)$ is non-zero. To arrive at a closed-form result, the integration domain is extended to the square delimited by dashed lines in the same figure. The factor $|H_c(f_2)|^2$ is shown at the top of the figure and seen to decay rapidly, even for a moderate value of $B\tau_c=20$. Contour lines of this function are overlaid and show that the function only overlaps a very small part of the extended integration domain, and in most of it, its value has substantially decayed.
The approximation will be more accurate the larger the bandwidth $B$ compared $f_c$ and the faster the integrand decays, i.e., the smaller $f_c\sim 1/\tau_c$. Hence the error is smaller the larger the product $B\tau_c$. 
The approximation is consistent with those made in the derivation of the integral form~\eqref{eq:gnlifinal} of the Gaussian noise model (see the discussion around~\eqref{eq:convatan} in Appendix~\ref{sec:app:gn} and the derivation of the closed form in Appendix~\ref{sec:app:closedformideal}. Both assume that $1/(B\tau_c)\ll 1$, which is typically fulfilled in wideband transmission systems. 
The extension of the integral domain therefore leads to only a small error, as will be validated below.

\begin{figure}[!t]
\centering
\includegraphics[width=\columnwidth]{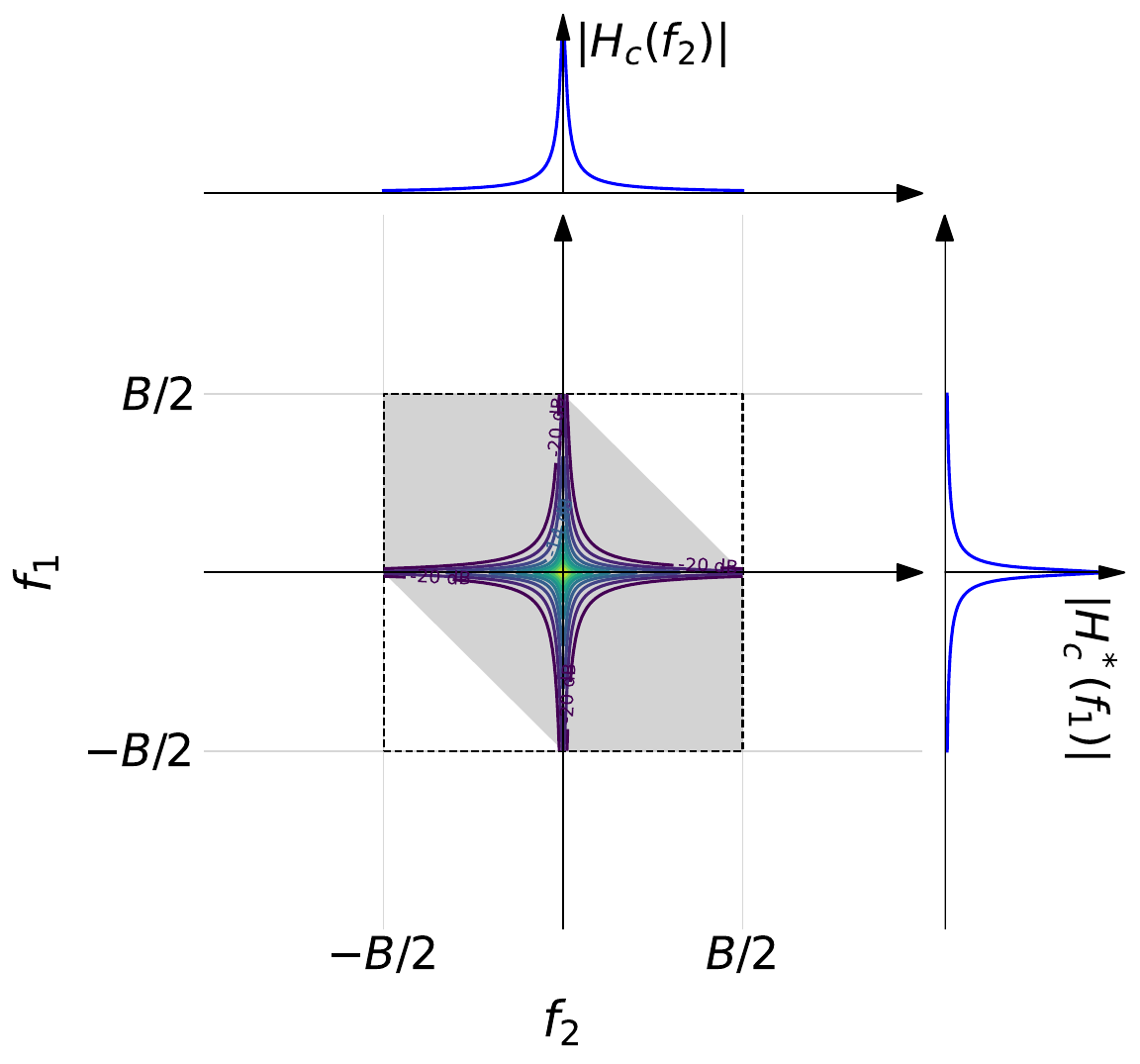}
\caption{Schematic representation of the integration region where the product $g_\text{in}(f_1)g_\text{in}(f_2)g_\text{in}(f_1+f_2)$ is non-zero (gray shaded area) and the extended integration region (dashed square) used in computation of the two-dimensional integral~\eqref{eq:gnlifinal} for the second contribution to the closed-form approximation proportional to 
$H_c(f_2)H_c^*(f_1)$. The magnitude of the functions $|H_c(f_2)|$ and $|H_c^*(f_1)|$ is shown as well as contour lines indicating the relative magnitude of their product ($\tau_c=100$ ps and $B\tau_c=20$).}
\label{fig:integralb}
\end{figure}

Note that since the filter function is independent of $f_1$, XCI contributions are added with equal weight independent of the spectral distance to the COI. As a result this term is dominant and has no analogue in fiber, where XCI decays with increasing frequency separation from the COI.
The second term depends on both frequencies $f_1$ and $f_2$. Its contour plot in Fig.~\ref{fig:integralb} has a hyperbolic shape reminiscent of the FWM efficiency that arises in the derivation of the GN model~\cite{Poggiolini2012}. Also here the integration region can be extended to the square while incurring a small error. Further details on the derivation of the closed-form approximation can be found in Appendix~\ref{sec:app:closedformideal}.

We are generally interested in the nonlinear $\text{NSR}$ for wideband WDM signals. Since edge effects are confined to a narrow band of width $\propto f_c$ at the band edges, the nonlinear $\text{NSR}$ 
for a channel of bandwidth $B_\text{ch}$ can be computed as $\text{NSR}_\text{NL} \approx G_\text{NL}(f_0)B_\text{ch}/P_\text{ch}$. Here $N_\text{ch}B_\text{ch}=B$ and $N_\text{ch}P_\text{ch}=P_\text{out}$. This leads to the following expression for $\text{NSR}_\text{NL}$ (see Appendix~\ref{sec:app:closedformideal}):
\begin{align}
\text{NSR}_\text{NL}\! =&\! \frac{1}{4}\!\left(1+\alpha_H^2\right)\! \frac{1}{1+\frac{P_\text{out}}{P_\text{sat}}}\left(\frac{P_\text{out}}{P_\text{sat}}\right)^2\!\left(1-\frac{1}{G}\right)^2\!\notag\\
&\times
\left[\frac{1}{2 B\tau_c}+\left(\frac{1}{2 B\tau_c}\right)^2\right].
\label{eq:nsrwdm1}
\end{align}
The first and second term in angular brackets in~\eqref{eq:nsrwdm1} stem from the corresponding terms in the third line of~\eqref{eq:gnlifinal}. From the previous discussion we expect the first term to be dominant, in particular for a wideband WDM signal dominated by XCI. Indeed, for a 6 THz signal and $\tau_c=100$ ps, $x=1/(2B\tau_c) = 8.3\cdot 10^{-4}$ and clearly $x^2\ll x$.
Even for a single channel of bandwidth 75 GHz, the second term is more than one order of magnitude (15 times) smaller. Neglecting the second term in angular brackets leads to the final simple result
\begin{align}
\text{NSR}_\text{NL}\! =\! \frac{1}{4}\!\left(1+\alpha_H^2\right)\! \frac{1}{1+\frac{P_\text{out}}{P_\text{sat}}}\left(\frac{P_\text{out}}{P_\text{sat}}\right)^2\!\left(1-\frac{1}{G}\right)^2\!
\frac{1}{2 B\tau_c}.
\label{eq:nsrwdm3}
\end{align}
For a single channel and above parameters, the difference between the two approximations is $10\log_{10}(1+1/15)$ or about 0.25 dB. Neglecting a factor $1 + 1/(2 B\tau_c)$ \emph{reduces} the NSR so that this simplification is not conservative. While care should be taken when applying the GN model to a single channel or narrowband signals, the error decreases further with increasing bandwidth.
A detailed analysis of the accuracy of the closed-form result is given in Section~\ref{sec:error}.

For comparison, the corresponding result for a single fiber span and ideal Nyquist-WDM over bandwidth $B$ (accounting for both polarizations) is given by~\cite{Poggiolini2012}
\begin{align}
\text{NSR}_\text{NL}^\text{fib}\! =\! \frac{8}{27}\gamma^2 P_\text{in}^2 L_\text{eff}^2 \frac{\text{asinh}(\frac{\pi^2}{2} |\beta_2|L_\text{eff}B^2)}{\pi |\beta_2|L_\text{eff}B^2},
\label{eq:nsrfib}
\end{align}
where $L_\text{eff}=(1-e^{-\alpha L})/\alpha$ is the effective length and $\alpha$, $\beta_2$ and $\gamma$ are the fiber attenuation, group velocity dispersion coefficient and nonlinearity parameter, respectively. For fixed power, $\text{NSR}_\text{NL}^\text{fib}$ asymptotically decays as $(1/B) \log(B)/B$ (using that $\text{asinh}(x)\approx \log(2x)$ for large $x$) and somewhat faster than $\text{NSR}_\text{NL}$ for an SOA ($\propto 1/B$).

\subsection{Numerical results for NSR of WDM signals}
\label{sec:wdmresults}

In this section we illustrate the dependence of the nonlinear NSR on the Agrawal model parameters and signal bandwidth. We compare the closed-form expression for WDM signals to numerical Agrawal model simulations. 
We first use a Gaussian distributed input signal with rectangular power spectral density of bandwidth $B=N_\text{ch}B_\text{ch}$, corresponding to the assumptions made in the closed-form GN-model derivation and as shown in Fig.~\ref{fig:spectrumideal}.

In order to estimate the nonlinear noise in the simulation, we compare the SOA output waveform to a reference signal without gain fluctuations, i.e., where $h(t)$ in~\eqref{eq:field} is replaced by the \emph{average} gain coefficient $\overline{h}$: $E^\text{ref}_\text{out}(t) = E_\text{in}(t)\exp\left((1-j\alpha_H)\overline{h}/2\right)$. The reference signal experiences an average gain compression and phase rotation. 
After filtering the COI with bandwidth $B_\text{ch}$ in the center of the band with a boxcar filter for both the Agrawal model output field and the reference field, we estimate the noise from the resulting samples via
\begin{equation}
P_\text{NL} = \frac{1}{N_s} \sum_{n=1}^{N_s} \left|s^\text{out}_n - s^\text{ref}_n\right|^2,
\label{eq:pnoise}
\end{equation}
which is simply an estimator of the variance based on the known mean. From $P_\text{NL}$ we compute the nonlinear NSR as $\text{NSR}_\text{NL}=P_\text{NL}/P_\text{ch}$.

\begin{figure}[t!]
\centering
\includegraphics[width=0.93\columnwidth]{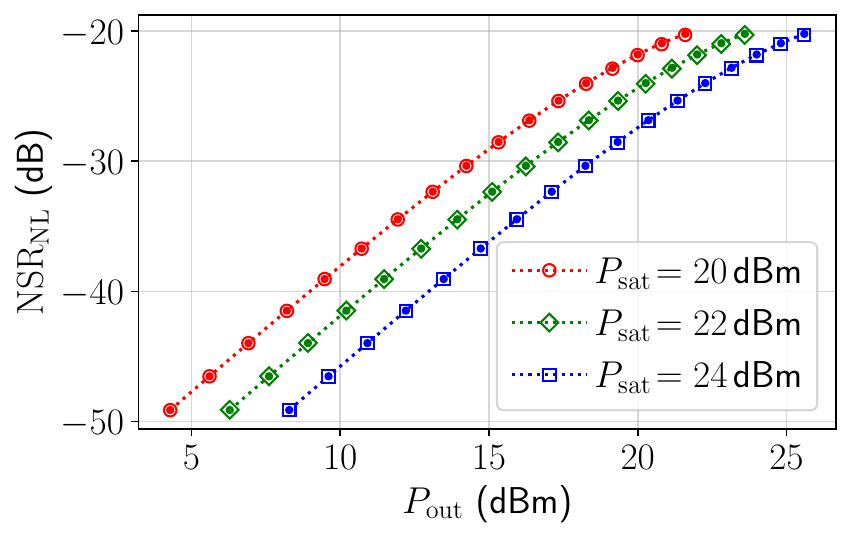}
\caption{Comparison of the nonlinear contribution to the NSR for the GN model (lines with small points) and numerical Agrawal model simulations (open symbols) for fixed $\tau_c=100$ ps, $\alpha_H=5$, $G_0=10$ dB and $N_\text{ch}=20$ (1.5 THz bandwidth), as a function of the total output power for different values of $P_\text{sat}$.}
\label{fig:nsrnlvspout_psat}
\includegraphics[width=0.93\columnwidth]{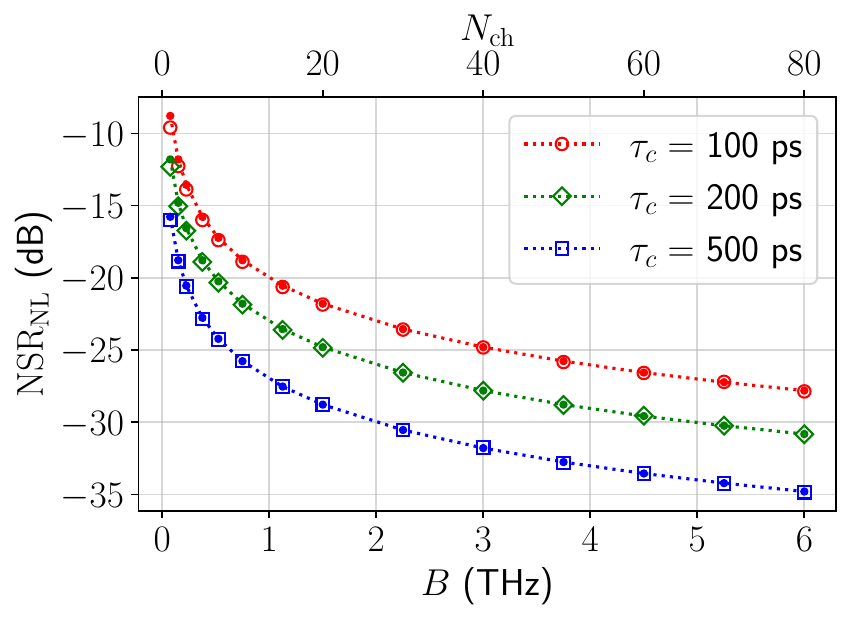}
\caption{Nonlinear contribution to the NSR for fixed $\alpha_H=5$, $G_0=10$ dB, $P_\text{sat}=24$ dBm and fixed total output power $P_\text{out}=P_\text{sat}$. The channel spacing is 75 GHz. $\text{NSR}_\text{NL}$ decreases with increasing bandwidth and increasing $\tau_c$.\\}
\label{fig:nsrnlvsb_fp_tauc}
\includegraphics[width=0.93\columnwidth]{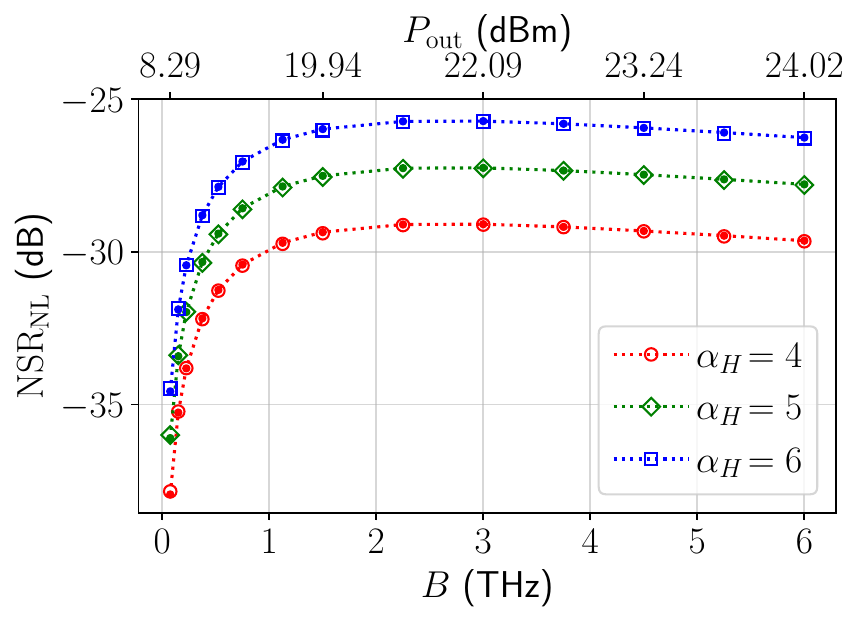}
\caption{Nonlinear contribution to the NSR for fixed $\tau_c=100$ ps, $P_\text{sat}=24$ dBm, $G_0=10$ dB and fixed per-channel power. The channel spacing is 75 GHz. $\text{NSR}_\text{NL}$ increases with increasing $\alpha_H$.}
\label{fig:nsrnlvsb_alpha}
\end{figure}

In Figs.~\ref{fig:nsrnlvspout_psat}-\ref{fig:nsrnlvsb_alpha} we illustrate its dependence on the Agrawal parameters $P_\text{sat}$, $\alpha_H$ and $\tau_c$.
We use a relatively small $G_0$ of 10 dB in all simulations to illustrate that the formula works in the regime where $1/G$ is not negligible compared to 1.

In Fig.~\ref{fig:nsrnlvspout_psat} we first compare~\eqref{eq:nsrwdm3} (dashed lines with dots) with simulations of the Agrawal model (open symbols) as a function of $P_\text{out}$.
Very good overall agreement is obtained, even for the largest values of $P_\text{out}$, which exceed the respective $P_\text{sat}$. According to~\eqref{eq:gaincurve}, the gain is compressed by a factor $\sim 1/e$ at $P_\text{out}=P_\text{sat}$, or more than 4 dB. This shows that the power dependence of the nonlinear $\text{NSR}$ is adequately captured by the model even in the strongly compressed regime.
As expected, $\text{NSR}_\text{NL}$ increases with increasing $P_\text{out}$ due to increasing nonlinear gain compression, but the increase of the nonlinear NSR slows down as the amplifier is increasingly saturated. The curves for different $P_\text{sat}$ are simply offset horizontally.
Equation~\eqref{eq:nsrwdm3} only depends on $P_\text{out}$ through the ratio $P_\text{out}/P_\text{sat}$. When plotted as a function of $P_\text{out}/P_\text{sat}$, the curves would collapse onto a single line.

Fig.~\ref{fig:nsrnlvsb_fp_tauc} shows $\text{NSR}_\text{NL}$ for fixed total power $P_\text{out}=P_\text{sat}$ as a function of bandwidth $B=N_\text{ch}B_\text{ch}$, or equivalently, the number of channels $N_\text{ch}$. $\text{NSR}_\text{NL}$ decreases because the low-pass filtered power fluctuations diminish with increasing bandwidth. The curve flattens off since the relative change in bandwidth decreases for increasing channel count. When the abscissa is plotted on a logarithmic scale, a straight line is obtained. 
Increasing the value of the carrier lifetime $\tau_c$ also has the effect of reducing power fluctuations and hence decreases $\text{NSR}_\text{NL}$.
Visible deviations between the formula and the numerical simulation are present only at small channel count (small bandwidth) and small $\tau_c$. Note that power $P_\text{out}=P_\text{sat}$ is artificially high for a few channels and therefore this regime is usually not relevant in practice. The error is mainly due to the approximations underlying the closed-form result and is analyzed in Section~\ref{sec:error}. 

In Fig.~\ref{fig:nsrnlvsb_alpha} we finally show $\text{NSR}_\text{NL}$ as a function of bandwidth $B$ or total output power $P_\text{out}$ when channel count increases for fixed per-channel power. Both bandwidth and power hence increase with $N_\text{ch}$. 
The per-channel power is chosen such that the total output power equals the saturation power for a fully loaded band (6 THz). $\text{NSR}_\text{NL}$ first sharply increases and then almost saturates. This is because of two competing effects shown in the previous two plots: (i) increasing the output power at fixed bandwidth increases nonlinear gain compression and hence gain and phase fluctuations, while (ii) increasing bandwidth at fixed power reduces power fluctuations and nonlinear distortions. Here both effects almost cancel. $\text{NSR}_\text{NL}$ is proportional to $B^{-1}\propto N_\text{ch}^{-1}$ and proportional to $P_\text{out}^2/(1+P_\text{out}/P_\text{sat})$. The latter grows significantly more slowly than $N_\text{ch}^2$ due to the combined effects of the denominator and gain compression at high power.
Finally, increasing the linewidth enhancement factor $\alpha_H$ is seen to increase NSR.  This is expected, since it increases the phase fluctuations according to~\eqref{eq:field}.

\subsection{Error analysis}
\label{sec:error}

\begin{figure}[!t]
\centering
\includegraphics[width=0.93\columnwidth]{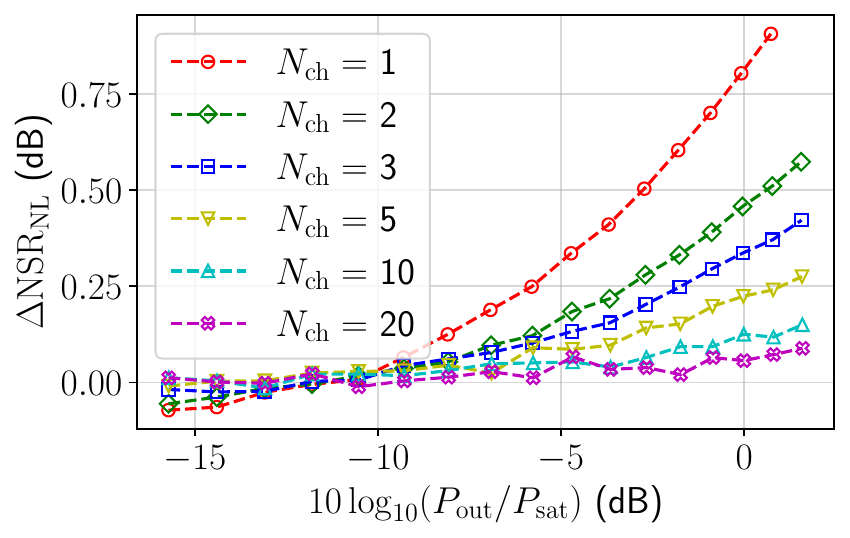}
\caption{NSR error of the closed-form GN model expression relative to numerical simulations of the Agrawal model as a function of total output power and for various channel counts. $P_\text{sat}=24 $ dBm, $G_0=10$ dB, $\tau_c=100$ ps, $\alpha_H=5$.}
\label{fig:errnsrnlvspch_nch_apprx3}
\end{figure}

\begin{figure}[!t]
\centering
\includegraphics[width=0.93\columnwidth]{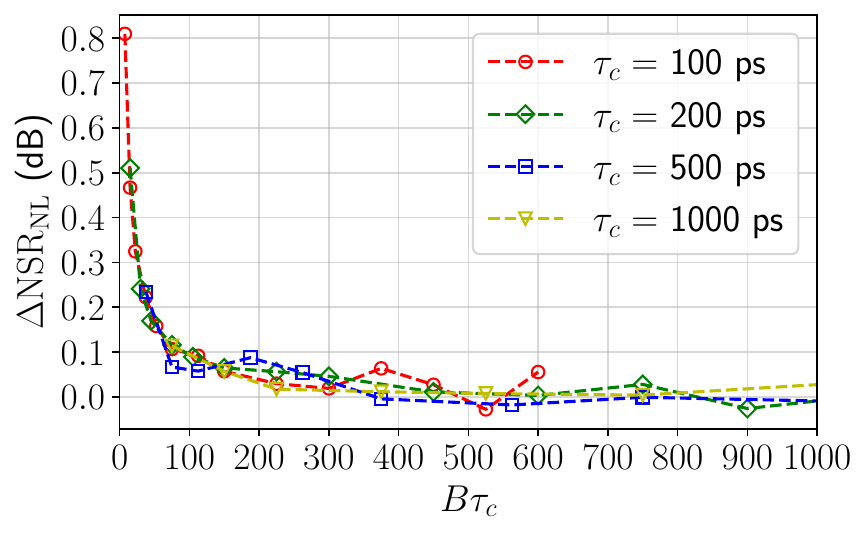}
\caption{NSR error of the closed-form GN model expression relative to numerical simulations of the Agrawal model, corresponding to Fig.~\ref{fig:nsrnlvsb_fp_tauc} for $P_\text{out}=P_\text{sat}$. For $B\tau_c\gtrsim 100$, the SNR error is of the order of 0.1 dB.}
\label{fig:errnsrnlvsbt_fp_tauc_apprx3}
\end{figure}

\begin{figure}[!ht]
\centering
\includegraphics[width=0.93\columnwidth]{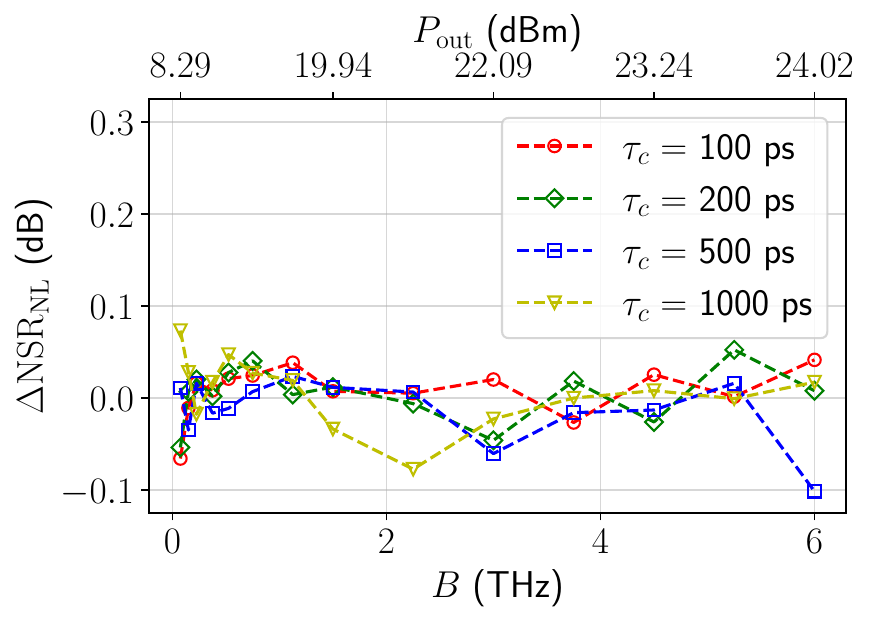}
\caption{NSR error of the closed-form GN model expression relative to numerical simulations of the Agrawal model, corresponding to Fig.~\ref{fig:nsrnlvsb_alpha} for $\alpha_H=5$.}
\label{fig:errnsrnlvsb_tauc_apprx3}
\end{figure}

In the following we assess the accuracy of the GN-model closed-form expression~\eqref{eq:nsrwdm3} in terms of the deviation relative to the simulation result, i.e.:
\begin{equation}
\Delta\text{NSR}_\text{NL} = 10\log_{10}(\text{NSR}_\text{NL}^\text{GN} / \text{NSR}_\text{NL}^\text{SIM})\ (\text{dB}).
\label{eq:deltansrdef}
\end{equation}
$\Delta\text{NSR}_\text{NL}$ is defined such that a positive value implies a conservative error, i.e., NSR is overestimated by the formula.

In the above results, visible deviations only occur in Fig.~\ref{fig:nsrnlvsb_fp_tauc} for high power and small channel count.
In Fig.~\ref{fig:errnsrnlvspch_nch_apprx3} we show the NSR error as a function of total output power relative to $P_\text{sat}$, for various channel counts. The error increases with power, as expected for a perturbation theory. It is slightly negative for very low power and largest for a single channel, for which it attains $\sim 0.8$ dB at $P_\text{out}=P_\text{sat}$. The error decreases rapidly with increasing channel count and is smaller than $0.1$ dB for $N_\text{ch}=20$ at $P_\text{out}=P_\text{sat}$.

To better understand the dependence of the error on signal bandwidth, we plot $\Delta\text{NSR}_\text{NL}$ corresponding to Fig.~\ref{fig:nsrnlvsb_fp_tauc}, i.e., for high fixed output power $P_\text{out}=P_\text{sat}$, in Fig.~\ref{fig:errnsrnlvsbt_fp_tauc_apprx3}.
As discussed above, the derivation of the GN-model involves the assumption $1/(2 B\tau_c)\ll 1$. We therefore plot the error as a function of the product $B\tau_c$, for different values of $\tau_c$. All curves collapse onto a single one, confirming that at fixed power, the error is only a function of $B\tau_c$. It decreases rapidly and converges to zero when $B\tau_c$ grows large. For $B\tau_c\gtrsim 100$ the error is smaller than $0.1$ dB.

In Fig.~\ref{fig:errnsrnlvsb_tauc_apprx3} we plot the error for the more relevant case of fixed per-channel power as a function of bandwidth. This plot corresponds to Fig.~\ref{fig:nsrnlvsb_alpha} for $\alpha_H=5$. The error is very small ($<0.1$ dB) for all band fillings and of the order of the statistical fluctuations due to the particular random realizations of the Gaussian signals.
The formula provides an accurate estimate of the nonlinear noise for typical application scenarios.

\section{Discussion}
\label{sec:discussion}

The closed-form expression~\eqref{eq:nsrwdm3} accurately captures the nonlinear contribution to the NSR from a standalone SOA.
It has been derived for a single SOA, and as such neglects the coherent accumulation of nonlinear noise that occurs in a transmission system.
In the GN model of fiber links, it is customary to describe the effect of coherent noise accumulation through a coherence correction factor $\epsilon_f$~\cite{Poggiolini2014}. It is defined through the scaling relation $G_\text{f}(N_\text{s})=N_\text{s}^{1+\epsilon_f}G_{\text{f}}(1)$, where $G_\text{f}(N_\text{s})$ is the nonlinear noise PSD of the link (we omit the subscript NLI for brevity), $G_{\text{f}}(1)$ is the nonlinear noise PSD of a single span and $N_\text{s}$ is the number of spans.

In a transmission system with SOAs, the noise PSD has contributions from the fibers (f), the SOAs (s) and their interference (sf), such that $G(N_s) = G_{\text{s}}(N_s) + G_{\text{f}}(N_s) +G_{\text{sf}}(N_s)$. Defining a coherence correction for each as $G_{\alpha}^\text{cc}=G_\alpha -G_{\alpha}^\text{inc}$, with $\alpha\in\{\text{s},\text{f},\text{sf}\}$, the corresponding exponents are defined as $\epsilon_\alpha=\ln(1+G_{\alpha}^\text{cc}/G_{\alpha}^\text{inc})/\ln(N_s)$.
As opposed to incoherent accumulation, for which the noise PSD scales linearly with $N_s$ as $G_\alpha^\text{inc}(N_s)=N_s G_{\alpha}(1)$, these exponents describe the coherent enhancement of the respective noise contribution: $G_\alpha(N_s)=N_s^{1+\epsilon_\alpha} G_{\alpha}(1)$. The combined effect of coherent noise accumulation can be captured through $\epsilon_\text{tot}$ defined through $G(N_\text{s}) = N_\text{s}^{1+\epsilon_\text{tot}}\sum_\alpha G_{\alpha}(1)$. The relation between $\epsilon_\text{tot}$ and the $\epsilon_\alpha$ is easily found. When the $\epsilon_\alpha$ are small, $\epsilon_\text{tot}$ is approximately given by the weighted sum $\sum_\alpha w_\alpha \epsilon_\alpha$, where the weights capture the relative strength of each contribution: $w_\alpha=G_{\alpha}(1)/\sum_\beta G_{\beta}(1)$.

The coherence correction factor $\epsilon_\text{f}$ is the one familiar from the GN model for fiber links.
When SOAs are separated by a sufficiently long fiber, noise correlations are often negligible~\cite{Hafermann2025} and $\epsilon_\text{s}$ will be very small. Note, however, that when $N_\text{soa}=2$ SOAs are directly cascaded (e.g. in a two-stage amplifier) and phase noise dominates, the nonlinear noise tends to accumulate coherently. This leads to a nonlinear noise power that is 3 dB higher ($\epsilon_\text{s}=\ln 2/\ln N_\text{soa})$ compared to the case of incoherent accumulation, unless the correlations are reduced by strong amplitude fluctuations, or large SOA parameter mismatch~\cite{Hafermann2025}.
The interaction between the SOA and fiber Kerr nonlinear distortions, that is captured by $\epsilon_{\text{sf}}$, can become relevant in long-haul transmission systems~\cite{Ghazisaeidi2019}. Depending on the sign of $\alpha_H$, it can be positive or negative. The simultaneous treatment of fiber and SOA nonlinear distortions appears to preclude a closed-form result. For highly accurate performance predictions where SOA fiber noise correlations cannot be neglected, the more complete theory of~\cite{Ghazisaeidi2019} should therefore be used.
In this case, our finding that a complete treatment gain compression increases nonlinear distortions by a factor $1+P_\text{out}/P_\text{sat}$ compared to first-order perturbation theory, should be taken into account.

The assumption of Gaussian signal statistics facilitates a closed-form result but neglects the dependence of the nonlinear noise on the modulation format, which is significant. 
It turns out that the closed-form result approximately still holds for different modulation formats if a modulation-dependent coefficient $c_\text{mod}$ is included as follows:
\begin{align}
\text{NSR}_\text{NL}\! =\! c_\text{mod}\frac{1}{4}\!\left(1+\alpha_H^2\right)\! \frac{1}{1+\frac{P_\text{out}}{P_\text{sat}}}\left(\frac{P_\text{out}}{P_\text{sat}}\right)^2\!\left(1-\frac{1}{G}\right)^2\!
\frac{1}{2 B\tau_c}.
\end{align}
The values of this coefficient can be tabulated for common modulation formats and as a function of entropy of shaped constellations~\cite{Hafermann2025b}.
The spectral shape of WDM channels can approximately be accounted for as outlined in Appendix~\ref{sec:app:closedformshaped}.

\section{Conclusions}
\label{sec:conclusions}

In this paper, a GN model of nonlinear distortions caused by a standalone SOA has been derived as described by the Agrawal model.
A physical interpretation of the general integral expression for the nonlinear noise PSD for a WDM signal has been given, along with accurate closed-form expressions. Their region of validity has been determined in numerical simulations. For ideal Nyquist-WDM, the formula is very accurate (with error $< 0.1$ dB) when the product of carrier lifetime and bandwidth $B\tau_c$ exceeds $\sim 100$.

The closed-form formula clearly exposes the dependence of the nonlinear noise on the Agrawal model parameters, the SOA output power and system bandwidth. It is useful to provide general intuition, or to guide SOA and system design, and to obtain a fast preliminary estimate of SOA nonlinear penalties in particular for large bandwidths where simulations are expensive and SOA--fiber nonlinear noise correlations can be neglected.

The GN-model has been derived for the idealized Agrawal model and under the assumption of Gaussian signal statistics. It turns out that these are not major limitations. In particular, it is possible to determine effective parameters such that the Agrawal model provides a quantitative description of real devices~\cite{Hafermann2024, Hafermann2025a}. This even remains true when the wavelength dependence of gain compression is taken into account. This is crucial in UWB systems~\cite{Zhao2023}, where the nonlinear penalty can exhibit a spread of as much as 3 dB in output power across C and L bands as observed experimentally~\cite{Demirtzioglou2023} and in simulation~\cite{Hafermann2024}.

The theory can further be extended, for example to describe the FWM efficiency for continuous-wave input (see Appendix~\ref{sec:app:fwmres}), to study the nonlinear noise in spectral gaps that can affect OSNR measurements, or to include the interaction of nonlinear noise from fibers and SOAs already captured in~\cite{Ghazisaeidi2019}. Our finding that higher-order terms stemming from nonlinear gain compression enhance nonlinear noise power by 3 dB at saturation can improve the accuracy of the result from first-order perturbation theory.

{\appendices

\section{Gain Compression}
\label{app:gaincompression}

In this Appendix we establish the relation governing the compression of the material gain coefficient $g(P)$ and provide an analytic formula for the Agrawal model gain curve.

The rate equation for the carrier density $N(z,t)$ underlying the Agrawal model~\cite{Agrawal1989} can be written in the form
\begin{align}
\frac{\partial N(z,t)}{\partial t} = \frac{I}{qV} - \frac{N(z,t)}{\tau_c} - \frac{N(z,t)-N_0}{\tau_c}\frac{P(z,t)}{P_\text{sat}}.
\label{eq:rateeq}
\end{align}
In the static case ($\partial N/\partial t=0$) and in absence of an input signal, the carrier density reaches its maximum value
\begin{align}
N^* = \frac{I\tau_c}{qV}.
\label{eq:agrawal_Nstar}
\end{align}
From the rate equation~\eqref{eq:rateeq} for a time-independent input signal it further follows that
\begin{align}
0 = N^* - N - (N-N_0)\frac{P}{P_\text{sat}}.
\end{align}
Solving for $N$ gives:
\begin{align}
N = \frac{N^* + N_0 P / P_\text{sat}}{1+P / P_\text{sat}}.
\label{eq:NofP}
\end{align}
Inserting this into the definition of the material gain coefficient yields:
\begin{align}
g(N) &\Let \Gamma a (N-N_0) 
=
\frac{\Gamma a}{1+P / P_\text{sat}}(N^*-N_0). 
\label{eq:agrawal_gstatic}
\end{align}
Since $N^*$ is the undepleted carrier density, $g(N^*)=\Gamma a(N^*-N_0)$ must equal the small-signal gain $g_0$, which can also be directly verified from its definition using~\eqref{eq:agrawal_Nstar} (see~\cite{Agrawal1989}): $g_0\Let\Gamma aN_0(I/I_0-1)$, where $I_0\Let qVN_0/\tau_c$ is the current required for transparency.
The compression of the material gain coefficient is hence described by
\begin{align}
g(P) = \frac{g_0}{1+P / P_\text{sat}}.
\label{eq:gcompression}
\end{align}
The Agrawal model gain curve~\eqref{eq:gaincompression} follows directly from this relation.
Using $\partial P/\partial z=(g-\alpha)P\approx gP$~\cite[Eq. (2.16)]{Agrawal1989} and $g(P)$ from above, separating variables and integrating yields
\begin{align}
\int_{P_\text{in}}^{P_\text{out}} dP \frac{1+P / P_\text{sat}}{P} = \int_0^L g_0 dz
\end{align}
or
\begin{align}
\log(P_\text{out}/P_\text{in}) + \frac{P_\text{out}}{P_\text{sat}} - \frac{P_\text{in}}{P_\text{sat}} = g_0 L.
\end{align}
Taking the exponential, defining $G_0 \Let \exp(g_0L)$ and $G\Let P_\text{out}/P_\text{in}$ 
yields an implicit relation that determines the Agrawal model gain curve:
\begin{align}
G = G_0\exp\left(-\Big(1-\frac{1}{G}\Big) \frac{P_\text{out}}{P_\text{sat}}\right).
\label{eq:gstatic}
\end{align}
For large gain, $1/G\ll 1$, we have $G(P_\text{out})\approx \exp(-P_\text{out}/P_\text{sat})$. 

The solution to~\eqref{eq:gstatic} can also be expressed analytically considering~\eqref{eq:hagrawal} in the static case ($dh/dt=0$):
\begin{align}
h = h_0 - \frac{P_\text{in}}{P_\text{sat}}[\exp(h)-1].
\label{eq:hagrawalstatic}
\end{align}
The solution to~\eqref{eq:hagrawalstatic} can be expressed as
\begin{equation}
h = h_0 + \frac{P_\text{in}}{P_\text{sat}} - W_0\left(\frac{P_\text{in}}{P_\text{sat}}\exp\left(h_0 + \frac{P_\text{in}}{P_\text{sat}}\right)\right),
\end{equation}
where $W_0(x)$ is the principal $(k=0)$ branch of the Lambert-$W$ function $W_k(x)$. It is the inverse function of $f(W)=W\exp(W)$ and hence known as the product logarithm.
Alternatively, \eqref{eq:hagrawalstatic} can be expressed in terms of $P_\text{out}$:
\begin{equation}
h = h_0 - \frac{P_\text{out}}{P_\text{sat}}[1 - \exp(-h)].
\label{eq:agrawalinstantaneouspout}
\end{equation}
Taking the exponential of this equation using $G=\exp(h)$ recovers~\eqref{eq:gstatic}. In this case, the solution is
\begin{equation}
h = h_0 - \frac{P_\text{out}}{P_\text{sat}} + W_0\left(\frac{P_\text{out}}{P_\text{sat}}\exp\left(-h_0 + \frac{P_\text{out}}{P_\text{sat}}\right)\right).
\label{eq:hagrawalinstantaneouspout}
\end{equation}
The solution $G(P_\text{out})$ to~\eqref{eq:gstatic} is obtained as $G = \exp(h)$.

\section{Heuristic derivation of power dependence of nonlinear NSR in first-order perturbation theory}
\label{sec:app:heuristic}

In this Appendix the dependence of the nonlinear noise-to-signal ratio on output power is derived based on a perturbative treatment up to first-order. The result shows that contrary to the main result~\eqref{eq:nsrwdm3} of this paper, the first-order result is smaller by a factor $1+P_\text{out}/P_\text{sat}$ and thus underestimates nonlinear noise. The origin of this discrepancy is gain compression. 
The correct result, which requires the summation of a perturbation series to infinite order, is derived in Appendix~\ref{sec:app:gn}.

In the following, time averages are denoted by a bar, e.g. $\overline{h}$, for clarity.
The gain fluctuations are described by Saleh's differential equation~\cite{Saleh1988}, which is equivalent to Agrawal's equation~\cite{Agrawal1989} and~\eqref{eq:hagrawal}:
\begin{align}
\frac{d h(t)}{d t} &= \frac{h_0-h(t)}{\tau_c} - \frac{1}{\tau_c}\frac{P_\text{out}(t)-P_\text{in}(t)}{P_\text{sat}}.
\label{eq:app:hsaleh}
\end{align}
Defining the gain fluctuations 
\begin{equation}
\Delta h(t) = h(t) - \overline{h}
\label{eq:deltagdef}
\end{equation}
around their average $\overline{h}$, the equation can be brought into the form
\begin{align}
\left(1+\tau_c\frac{d}{d t}\right) \Delta h(t) &= h_0 -\overline{h} - \frac{P_\text{out}(t)-P_\text{in}(t)}{P_\text{sat}}.
\label{eq:app:deltahdynamics}
\end{align}
For time-independent input $P_\text{in}(t)=\overline{P}_\text{in}$, the derivative $dh/dt$ vanishes and one obtains~\eqref{eq:hagrawalstatic}:
\begin{align}
0 &= h_0 -\overline{h} - \frac{\overline{P}_\text{out}-\overline{P}_\text{in}}{P_\text{sat}}.
\label{eq:app:hstatic}
\end{align}
Using the definition~\eqref{eq:deltagdef}, the output power can approximately be expressed as
\begin{align}
P_\text{out}(t) &= P_\text{in}(t)\exp(h(t)) = P_\text{in}(t)\exp(\overline{h})\exp(\Delta h(t))\notag\\
								&\approx \overline{P}_\text{in}\exp(\overline{h})(1+\Delta h(t)) + \Delta P_\text{in}(t)\exp(\overline{h})\notag\\
								&=\overline{P}_\text{out}\Delta h(t) + P_\text{in}(t)\overline{G}.
\label{eq:app:poutt}
\end{align}
Here the exponential $\exp(\Delta h(t))\approx 1+\Delta h(t)$ has been expanded to first order, and a term of order $\mathcal{O}(\Delta P_\text{in}(t)\Delta h(t))$ has been neglected. In the last line it has been used that $\overline{P}_\text{out}=\overline{P}_\text{in}\overline{G}$ and that $\exp(\overline{h})=\overline{G}$ is the average, time-independent gain.
The term proportional to $\Delta h(t)$ describes the nonlinear gain modulation of $P_\text{out}(t)$. This term is responsible for a renormalization of the recovery time.
Inserting this result into~\eqref{eq:app:deltahdynamics}, one obtains
\begin{align}
\left(1+\frac{\overline{P}_\text{out}}{P_\text{sat}}+\tau_c\frac{d}{d t}\right) \Delta h(t) = h_0 -\overline{h} &- \frac{P_\text{in}(t)(\overline{G}-1)}{P_\text{sat}}.
\end{align}
Defining the input power fluctuations 
\begin{equation}
\Delta P_\text{in}(t) = P_\text{in}(t) - \overline{P}_\text{in}
\label{eq:app:dpindef}
\end{equation}
and using the static solution~\eqref{eq:app:hstatic}, the time-independent terms cancel and one obtains
\begin{align*}
\left(1+\frac{\overline{P}_\text{out}}{P_\text{sat}}+\tau_c\frac{d}{dt}\right) \Delta h(t) = 
- \frac{\Delta P_\text{in}(t)\left(\overline{G} - 1\right)}{P_\text{sat}}.
\end{align*}
With the effective recovery time
$\tau_\text{eff} \Let \tau_c/(1 + \overline{P}_\text{out}/P_\text{sat})$, this becomes:
\begin{align*}
\left(1+\tau_\text{eff}\frac{d}{dt}\right) \Delta h(t) = 
- \frac{1}{1 + \overline{P}_\text{out}/P_\text{sat}}\frac{\Delta P_\text{in}(t)\left(\overline{G} - 1\right)}{P_\text{sat}}.
\end{align*}
By denoting the low-pass filter operation in time-domain as $F_\text{eff}[\cdot]$, this can symbolically be expressed as
\begin{align}
\Delta h(t) = - \frac{1}{1 + \overline{P}_\text{out}/P_\text{sat}}\frac{F_\text{eff}[\Delta P_\text{in}(t)]\left(\overline{G} - 1\right)}{P_\text{sat}}.
\label{eq:app:deltahlowpass}
\end{align}
Using~\eqref{eq:app:dpindef}, this can be rewritten in the form:
\begin{align}
\Delta h(t) = - \frac{1}{1 + \overline{P}_\text{out}/P_\text{sat}}F_\text{eff}\left[\frac{P_\text{in}(t)-\overline{P}_\text{in}}{\overline{P}_\text{in}}\right]\frac{\overline{P}_\text{out}}{P_\text{sat}}\left(1 - \frac{1}{\overline{G}}\right).
\label{eq:app:deltahfinal}
\end{align}
Note that the result of the filtering operation, $F_\text{eff}[P_\text{in}(t)/\overline{P}_\text{in}-1]$, is independent of input power.

It remains to express the nonlinear $\text{NSR}$ in terms of the gain fluctuations $\Delta h(t)$.
The equation for the electric field at SOA output is given by~\cite{Agrawal1989}:
\begin{equation}
E_\text{out}(t) = E_\text{in}(t)e^{\frac{1}{2}(1-j\alpha_H)h(t)}.
\label{eq:app:field}
\end{equation}
The dynamic gain $h(t)$ is split into its average and fluctuations: $h(t) = \overline{h} +\Delta h(t)$, so that
\begin{equation}
E_\text{out}(t) = E_\text{in}(t)e^{\frac{1}{2}(1-j\alpha_H)\overline{h}}e^{\frac{1}{2}(1-j\alpha_H)\Delta h(t)}.
\end{equation}
The nonlinear noise power is defined as the average of the absolute square of the fluctuations of the output field relative to the output field in absence of gain fluctuations (i.e., $\Delta h(t)\equiv 0$):
\begin{align}
P^{(1)}_\text{NL}&=\langle|E_\text{out}(t)-E_\text{in}(t)e^{(1-j\alpha_H)\overline{h}/2}|^2\rangle\notag\\
&= \langle|E_\text{in}(t)e^{(1-j\alpha_H)\overline{h}/2}(1-j\alpha_H)\Delta h(t)/2|^2\rangle\notag\\
&= \frac{1}{4}(1+\alpha_H^2)\overline{G}\langle|E_\text{in}(t)|^2\Delta h(t)^2\rangle\notag\\
&= \frac{1}{4}(1+\alpha_H^2)\overline{P}_\text{out}\langle\Delta h(t)^2\rangle.
\label{eq:app:pnl}
\end{align}
Here the superscript $(1)$ emphasizes that the result is valid up to first order, $\langle\cdot\rangle$ denotes expectation and $\overline{G}=\exp(\overline{h})$ as before. Here it has been used that $\Delta h(t)$, which is proportional to the low-pass filtered signal according to~\eqref{eq:app:deltahfinal}, is virtually uncorrelated with the input power fluctuations: $\langle|E_\text{in}(t)|^2\Delta h(t)^2\rangle\approx \langle|E_\text{in}(t)|^2\rangle\langle\Delta h(t)^2\rangle$.

The nonlinear $\text{NSR}$ is defined as $\text{NSR}_\text{NL}=P_\text{NL}/\overline{P}_\text{out}$. Combining~\eqref{eq:app:pnl} and~\eqref{eq:app:deltahfinal}, we finally obtain the $\text{NSR}_\text{NL}^{(1)}$ in first-order approximation:
\begin{align}
\text{NSR}^{(1)}_\text{NL}=& \frac{1}{4}\left(1+\alpha_H^2\right) \frac{1}{\left(1+\frac{P_\text{out}}{P_\text{sat}}\right)^2}\left(\frac{P_\text{out}}{P_\text{sat}}\right)^2\left(1-\frac{1}{\overline{G}}\right)^2 \times\notag\\
& \times \langle
F_\text{eff}[P_\text{in}(t)/\overline{P}_\text{in}-1]^2\rangle.
\label{eq:app:nsrheuristic}
\end{align}
The expectation of the filtered relative power fluctuations in the second line is independent of power and only depends on the normalized signal statistics. Comparing with the main result~\eqref{eq:nsrwdm1} of this paper, Eq.~\eqref{eq:app:nsrheuristic} \emph{underestimates} the nonlinear noise-to-signal ratio by a factor $1+P_\text{out}/P_\text{sat}$ and by as much as 3 dB when the SOA is operated at saturation. 
The power dependence of this result is consistent with the first-order perturbation theory of Ref.~\cite{Ghazisaeidi2019}. 
To fully account for gain compression, it is necessary to perform a partial summation of the deterministic terms arising in the perturbation series at all orders, which is accomplished in the next section.

\section{Derivation of the GN-Model}
\label{sec:app:gn}

In this section, the GN-theory for the Agrawal model is derived. 
The derivation roughly follows~\cite{Poggiolini2014detailed} which provides a detailed derivation of the GN model for the NLSE. Important differences are pointed out where adequate.

The following Fourier transform pair is used:
\begin{align}
\mathcal{F}[f(t)](f) &\Let \int_{-\infty}^{+\infty} dt\, f(t)\, e^{-j2\pi f t}\notag\\
\mathcal{F}^{-1}[\hat{f}(f)](t) &\Let \int_{-\infty}^{+\infty} df\, \hat{f}(f)\, e^{j2\pi f t}.
\label{eq:ftdef}
\end{align}

The equation governing the field in the Agrawal model is given by~\cite{Agrawal1989}:
\begin{align}
\frac{\partial E(z,t)}{\partial z} =& \frac{1}{2}\left(g(z,t)-\alpha\right)E(z,t) - j\frac{1}{2}\alpha_H g(z,t) E(z,t),\notag\\
=& \frac{1}{2}\left[(1-j\alpha_H)g_0-\alpha \right]E(z,t)\notag\\
 &+\frac{1}{2}(1-j\alpha_H)\Delta g(z,t)E(z,t),
\label{eq:agrawaldedz2}
\end{align}
where the gain fluctuation $\Delta g(z,t) \Let g(z,t)-g_0$ was defined. The gain dynamics are described by
\begin{align}
\frac{\partial g(z,t)}{\partial t} 
&=\frac{g_0-g(z,t)}{\tau_c} - \frac{1}{\tau_c}\frac{g(z,t)P(z,t)}{P_\text{sat}}.
\end{align}
To simplify notation, define the perturbation parameter $\epsilon \Let P_\text{sat}^{-1}$ and write
\begin{align}
\frac{\partial \Delta g(z,t)}{\partial t} 
&=-\frac{\Delta g(z,t)}{\tau_c} - \epsilon\frac{1}{\tau_c}g(z,t)P(z,t),
\end{align}
or equivalently
\begin{align}
\left(1+\tau_c\frac{\partial}{\partial t}\right)\Delta g(z, t)
&=-\epsilon g(z,t)|E(z,t)|^2.
\label{eq:deltag2}
\end{align}
Using that the Fourier transform of $E^*(t)$ is $E^*(-f)$ and the convolution theorem
\begin{align}
\mathcal{F}[E_1(t) E_2(t)](f) = \int_{-\infty}^{+\infty}df_1 E_1(f_1) E_2(f-f_1)
\label{eq:conv}
\end{align}
repeatedly, it is straightforward to show that
\begin{align}
\Delta g(z, f) = -&\epsilon \mathcal{F}[|E(z,t)|^2 g(z,t)](f)\, H_c(f)\notag\\
= -&\epsilon
\int_{-\infty}^{+\infty}df_1 \int_{-\infty}^{+\infty}df_2 E(z,f_1) E^*(z,f_1-f_2)\notag\\
&\qquad\qquad\qquad\quad\times g(z,f-f_2)\, H_c(f)\label{eq:deltagf2},
\end{align}
where $H_c(f) = 1/(1 + 2\pi i \tau_c f)$.
In the following, we abbreviate $\int df_1\equiv\int_{-\infty}^{+\infty}df_1$.
Using $g(z,f-f_2) = g_0\delta(f-f_2) + \Delta g(z,f-f_2)$ and repeatedly inserting the left-hand side into the right-hand side, the equation can be iterated:
\begin{align}
\Delta g(z,f) = &-\epsilon g_0
\int df_1 E(z,f_1) E^*(z,f_1-f) H_c(f)+\notag\\
& \epsilon^2 g_0 \int\! df_1\int\! df_2\int\! df_3 E(z,f_1) E^*(z,f_1-f_2)\times\notag\\
&E(z,f_3)E^*(z,f_3\!-\! f\!+\! f_2) H_c(f-f_2) H_c(f) \mp\ldots
\end{align}
Using the convolution theorem once more, one obtains
\begin{align}
&\mathcal{F}[\Delta g(z,t)E(z,t)](f) =\notag\\
 = -&\epsilon g_0
\int\! df_1 \int\! df_2 E(z,f_1) E^*(z,f_1\! -\! f_2) H_c(f_2) E(z,f\! - \! f_2)\notag\\
+ &\epsilon^2 g_0 \int\! df_1\ldots df_4 E(z,f_1) E^*(z,f_1\! -\! f_2) E(z,f_3) \times\notag\\
& E^*(z,f_3\! -\! f_4\! +\! f_2) H_c(f_4\! -\! f_2) H_c(f_4) E(z,f\! -\! f_4)\mp\ldots
\label{eq:secondorder}
\end{align}
Taking the Fourier transform of~\eqref{eq:agrawaldedz2} and inserting the previous result up to first order yields
\begin{equation}
\frac{\partial}{\partial z}E(z,f) = \hat{\Gamma}(z)E(z,f) + Q_\text{NLI}[E(z,f)],
\label{eq:dedzfull}
\end{equation}
where
\begin{align}
Q_\text{NLI}[E(z,f)] = -&\frac{1}{2}\left(1-j\alpha_H\right) 
\epsilon g_0 \int df_1 \int df_2 E(z,f_1) \times \notag\\
& E^*(z,f_1-f_2)H_c(f_2) E(z,f-f_2)
\end{align}
and 
\begin{align}
\hat{\Gamma}(z) = \frac{1}{2}\left[(1-j\alpha_H)g_0 -\alpha\right].
\label{eq:gammadef}
\end{align}
The corresponding linear equation is
\begin{align}
\frac{\partial}{\partial z}E(z,f) = \hat{\Gamma}(z)E(z,f),
\label{eq:elin1} 
\end{align}
whose solution is obtained by separation of variables, i.e., $dE/E=\hat{\Gamma}(z) dz$. Integrating both sides, one obtains
\begin{align}
\log(E(z,f)/E(0,f)) = \int_0^L dz\hat{\Gamma}(z) \teL \Gamma(z)
\label{eq:sepvar}
\end{align}
or
\begin{equation}
E_\text{LIN}(z,f) = e^{\Gamma(z)} E(0,f) \equiv e^{\Gamma(z)}E_\text{in}(f).
\label{eq:elin}
\end{equation}

$Q_\text{NLI}$ above is linear in $\epsilon$.
In order to treat the nonlinear term $Q_\text{NLI}[E(z,f)]$ in first-order perturbation theory in $\epsilon$, the field in $Q_\text{NLI}(z,f)$ is to be replaced by the zeroth-order approximation, i.e., by the field $E_\text{LIN}(z,f)$ in~\eqref{eq:elin}.
Inserting $E_\text{LIN}(z,f)$ into $Q_\text{NLI}[E(z,f)]$ yields:
\begin{align}
Q_\text{NLI}[E_\text{LIN}(z,f)] =
-&\frac{1}{2}\left(1-j\alpha_H\right) 
\epsilon g_0 \int\! df_1 \int\! df_2 E_\text{in}(f_1)\times \notag\\
&E_\text{in}^*(f_1\! -\! f_2) H_c(f_2) E_\text{in}(f\! -\! f_2)\times\notag\\
& e^{\Gamma(z) + \Gamma^*(z) + \Gamma(z)}.
\label{eq:qnli}
\end{align}

Before proceeding, recall that the signal model in the GN theory is
\begin{align}
E_\text{in}(f) &= \sqrt{\Delta f}\sqrt{G_\text{in}(f)}\sum_{n=-\infty}^\infty \xi_n \delta(f-n\Delta f),
\label{eq:signalmodel}
\end{align}
where $\xi_n$ are complex Gaussian random variables with zero mean and unit variance, such that $\langle \xi_m\xi_n^*\rangle=\delta_{mn}$, where $\langle\cdot\rangle$ denotes the expectation (see~\cite{Poggiolini2014detailed} for details).
To simplify notation, $\sum_n\Let\sum_{n=-\infty}^\infty$ is used. 
The power of a single random realization of the field is a random variable
\begin{align}
\tilde{P}_E &= \int |E_\text{in}(f)|^2 df=\sum_n |\xi_n|^2G_\text{in}(n\Delta f)\Delta f,
\label{eq:ptildee}
\end{align}
which in the limit $\Delta f\to 0$ is the sum over an increasingly large number of terms. It hence statistically converges to the ensemble average~\cite{Poggiolini2014detailed}
\begin{align}
P_E &= \int_{-\infty}^\infty G_E(f) df = \sum_n G_\text{in}(n\Delta f)\Delta f
\approx\int_{-\infty}^\infty G_\text{in}(f)df,
\label{eq:ptx}
\end{align}
which equals $P_\text{in}$.
The approximation becomes asymptotically exact in the limit $\Delta f\to 0$.
Inserting~\eqref{eq:signalmodel} into~\eqref{eq:qnli}, one obtains
\begin{align}
&Q_\text{NLI}[E_\text{LIN}(z,f)] = -\frac{1}{2}\left(1-j\alpha_H\right) 
\epsilon g_0 (\Delta f)^{\frac{3}{2}}\sum_{m,n,k} \xi_m \xi_n^* \xi_k\,\notag\\ 
&\times\int df_1 \int df_2 \sqrt{G_\text{in}(f_1)G_\text{in}(f_1-f_2)G_\text{in}(f-f_2)}H_c(f_2)\notag\\
&\quad \times e^{\Gamma(z) + \Gamma^*(z) + \Gamma(z)}\notag\\
&\quad\times \delta(f_1-m\Delta f)\delta(f_1-f_2-n\Delta f)\delta(f-f_2-k\Delta f).
\label{eq:qnlidef}
\end{align}
When the integrals are performed, the delta distributions require that $f_1 = m\Delta f$ and $f_2 = f_1-n\Delta f$. Hence $\delta(f-f_2-k\Delta f)=\delta(f-[m-n+k]\Delta f)$. 
Defining a new index $i$, the triple summation over $m,n,k$ can be regrouped into a summation over $i$ and a restricted triple sum over $m,n,k$ such that $m-n+k=i$. To this end, define the set
\begin{equation}
A_i \Let \{(m,n,k): m-n+k=i\}.
\label{eq:aidef}
\end{equation}
The set $A_i$ can be further split into two disjoint subsets, which require separate treatment, namely
\begin{equation}
X_i \Let \{(m,n,k): [m-n+k=i]\, \text{and}\, [m=n\, \text{or}\, k=n] \}
\end{equation}
and the coset of $X_i$ in $A_i$: $\tilde{A}_i \Let A_i \setminus X_i$. 
Symbolically, omitting the summands, this allows one to write
\begin{equation}
\sum_m\sum_n\sum_k=\sum_i\sum_{m,n,k\in A_i}=\sum_i\Big(\sum_{m,n,k\in X_i} + \sum_{m,n,k\in \tilde{A}_i}\Big)
\end{equation}
and to split the nonlinear interference term into two parts:
\begin{equation}
Q_\text{NLI}[E_\text{LIN}(z,f)] = Q_{\text{NLI},X_i}[E_\text{LIN}(z,f)] + Q_{\text{NLI},\tilde{A}_i}[E_\text{LIN}(z,f)].
\label{eq:qnli1}
\end{equation}
Consider the contribution $Q_{\text{NLI},X_i}[E_\text{LIN}(z,f)]$ first.
The sum over all triplets in $X_i$ contains three types of terms, which form three disjoint subsets. Symbolically the sum can hence be decomposed into three parts:
$$
\sum_i\sum_{m,n,k\in X_i} = \sum_k\sum_{m=n\neq k} + \sum_m\sum_{m\neq n=k} + \sum_{m = k = n} 
$$
Recalling that $i=m-n+k$, the following conditions hold:
\begin{align}
m = n \neq k & \implies i = k,\notag\\
m \neq n = k & \implies i = m,\notag\\
m = n = k & \implies i = m = n = k.
\end{align}
Therefore, the sums can be written 
\begin{align}
\sum_i\sum_{m,n,k\in X_i} = \sum_i\sum_{m=n\neq i} + \sum_i\sum_{n=k\neq i} + \sum_i 
\end{align}
For fixed $i$ the last sum on the right only contains a single term. In the limit $\Delta f\to 0$, $\sum_i$ hence becomes irrelevant compared to the sums over $m$ or $n$. On the other hand, each of the sums over $m$ or $n$ on the right are missing this term. Since one is free to add it in the limit $\Delta f\to 0$, these sums can be made unrestricted with respect to $i$:
\begin{align}
\sum_i\sum_{m,n,k\in X_i} = \sum_i\sum_{k=n} + \sum_i\sum_{m=n}
\end{align}
With~\eqref{eq:qnlidef}, $Q_{\text{NLI},X_i}[E_\text{LIN}(z,f)]$ in \eqref{eq:qnli1} is as follows:
\begin{align}
\!\!\!Q_{\text{NLI},X_i}&[E_\text{LIN}(z,f)]\! =\!
-\frac{1}{2}\left(1\!-\!j\alpha_H\right)\epsilon g_0 (\Delta f)^{\frac{3}{2}}\sum_i \delta(f-i\Delta f)\notag\\ 
&\!\!\!\sum_{m,n,k\in X_i}\!\!\! \xi_m \xi_n^* \xi_k \sqrt{G_\text{in}(m\Delta f)G_\text{in}(n\Delta f)G_\text{in}(k\Delta f)}\notag\\
&\times H_c([m-n]\Delta f) e^{\Gamma(z) + \Gamma^*(z) + \Gamma(z)}
\notag\\
=-&\frac{1}{2}\left(1-j\alpha_H\right) 
\epsilon g_0 \sqrt{\Delta f} \sqrt{G_\text{in}(f)}\sum_i \delta(f-i\Delta f)\xi_i e^{\Gamma(z)}\notag\\
&\times\left(\sum_{k=n} |\xi_k|^2 \Delta f G_\text{in}(k\Delta f)H_c(f-k\Delta f)e^{2\text{Re} \Gamma(z)}\right.\notag\\
&\quad\left.+\sum_{m=n} |\xi_m|^2 \Delta f G_\text{in}(m\Delta f)e^{2\text{Re}\Gamma(z)}\right).
\label{eq:qnlixi}
\end{align}
The above can be simplified using the definition of the signal model \eqref{eq:signalmodel}, as well as the definition of the linear field~\eqref{eq:elin}. 
By the same arguments as above, a given random realization of $Q_{\text{NLI},X_i}$ can be replaced by its ensemble average~\cite{Poggiolini2014detailed}. Taking the expectation using $\langle|\xi_m|^2\rangle=1$ and performing the continuum limit $\Delta f \to 0$, one obtains
\begin{align}
Q_{\text{NLI},X_i}[E_\text{LIN}&(z,f)]=-\frac{1}{2}\left(1-j\alpha_H\right) 
\epsilon g_0 E_\text{LIN}(z,f) e^{2\text{Re} \Gamma(z)}\notag\\
\times& \left( \int df' G_\text{in}(f')H_c(f-f') + \int df' G_\text{in}(f')\right).
\label{eq:qnlix}
\end{align}
The convolution term in the second line can be neglected compared to the second. Consider, for simplicity, a signal with a rectangular power spectral density that is constant and non-zero within the bandwidth $B$ and zero outside. Without loss of generality, choose $f=0$ as the center of the band:
\begin{align}
&\int df' G_\text{in}(f')H_c(f-f') = G_\text{in}\int_{-B/2}^{B/2} H_c(-f')\,df'\notag\\
&= G_\text{in} \int_{-B/2}^{B/2} \frac{1}{1-j2\pi\tau_c f'}\,df'= G_\text{in}\frac{\arctan(\pi B\tau_c)}{\pi\tau_c}.
\label{eq:convatan}
\end{align}
For a carrier lifetime of $\tau_c\sim 100$ ps, or $\tau_c^{-1}=10$ GHz, and a bandwidth of 100 GHz, $B\tau_c=10$. Using 
\begin{align}
\arctan(\pi B\tau_c)\approx\pi/2
\label{eq:approxatan}
\end{align}
(within a relative error $<$ 0.1 dB), the convolution equals $G_\text{in}/(2\tau_c)$. It is smaller than $\int df' G_\text{in}(f')=BG_\text{in}=P_\text{in}$ by a factor $1/(2B\tau_c)=1/20$ and will be neglected.

Using the definition~\eqref{eq:sepvar} of $\Gamma(z)$ and~\eqref{eq:gammadef}, one has $2\text{Re}\Gamma(z) = (g_0-\alpha)z$. Equation~\eqref{eq:qnlix} then simplifies to
\begin{align}
Q_{\text{NLI},X_i}[E_\text{LIN}(z,f)]=&\!-\!\frac{1}{2}\!\left(1\!-\!j\alpha_H\right)\!
g_0 E_\text{LIN}(z,f) \epsilon P_\text{in}e^{(g_0-\alpha)z},
\label{eq:qnlixsimp}
\end{align}
which is linear in the field $E_\text{LIN}(z,f)$.
Coming back to the differential equation~\eqref{eq:dedzfull} for the field, using the decomposition~\eqref{eq:qnli1} and the above result~\eqref{eq:qnlixsimp}, one can write
\begin{equation}
\frac{\partial}{\partial z}E(z,f) = \hat{\tilde{\Gamma}}(z)E(z,f) + Q_{\text{NLI},\tilde{A}_i},
\label{eq:dedzfull2}
\end{equation}
where now~\eqref{eq:qnlixsimp} has been absorbed into the definition of $\hat{\tilde{\Gamma}}$:
\begin{equation}
\hat{\tilde{\Gamma}}(z) \Let \frac{1}{2}\left[(1-j\alpha_H)g_0(1 - \epsilon P_\text{in} e^{(g_0-\alpha)z}) -\alpha\right].
\label{eq:gammahattildedef}
\end{equation}

A key assumption of the GN model derivation is that the remaining nonlinear term $Q_{\text{NLI},\tilde{A}_i}[E(z,f)]$ in~\eqref{eq:dedzfull2} acts as a pure source term. I.e., it is actually assumed to be independent of $E(z,f)$. This assumption has been justified by the fact that due to the double convolution in \eqref{eq:qnli}, 
$Q_{\text{NLI},\tilde{A}_i}[E(z,f)]$ contains contributions from products of fields at many different frequencies, which are treated as random variables.
Hence $Q_{\text{NLI},\tilde{A}_i}[E(z,f)]$ is regarded as a random variable that is independent of $E(z,f)$. This argument does not hold for $Q_{\text{NLI},X_i}[E(z,f)]$, which is why it is treated separately.

With this assumption, the propagation equation can be written
\begin{equation}
\frac{\partial}{\partial z}E(z,f) = \hat{\tilde{\Gamma}}(z)E(z,f) + Q_{\text{NLI},\tilde{A}_i}(z,f).
\end{equation}
where here the dependence of $Q_{\text{NLI},\tilde{A}_i}$ on the field is omitted, as per the argument above.
This equation can be solved by introducing an integrating factor $M(z,f)$:
\begin{align}
M(z,f) Q_{\text{NLI},\tilde{A}_i}\!\! = &M(z,f)\left(\frac{\partial}{\partial z}E(z,f) - \hat{\tilde{\Gamma}}(z)E(z,f)\right).
\end{align}
In order to write the right-hand side as a total differential
\begin{align}
\frac{d}{dz} [M(z,f)E(z,f)] =& M(z,f)\frac{\partial E(z,f)}{\partial z}
\!+\! E(z,f)\frac{\partial M(z,f)}{\partial z},
\end{align}
one must have
\begin{equation}
\frac{\partial}{\partial z}M(z,f) = - M(z,f) \hat{\tilde{\Gamma}}(z).
\label{eq:mzf}
\end{equation}
Separation of variables and integration similarly to~\eqref{eq:sepvar} yields
\begin{equation}
M(z,f) = e^{-\int_0^z dz' \hat{\tilde{\Gamma}}(z')} = e^{-\tilde{\Gamma}(z)},
\label{eq:mdef}
\end{equation}
where, without loss of generality, the integration constant is set to zero, so that $M(0,f)=1$. Equation~\eqref{eq:mzf} also implies that
\begin{equation}
M(z,f) Q_{\text{NLI},\tilde{A}_i}(z,f) = \frac{d}{dz} [M(z,f)E(z,f)].
\end{equation}
Hence
\begin{equation}
M(z',f)E(z',f)\Big|_0^z = \int_0^z dz' M(z',f) Q_{\text{NLI},\tilde{A}_i}(z',f)
\end{equation}
or
\begin{equation}
M(z,f)E(z,f) = E(0,f) + \int_0^z dz' M(z',f) Q_{\text{NLI},\tilde{A}_i}(z',f).
\end{equation}
Inserting the above result~\eqref{eq:mdef} for $M$ and solving for $E(z,f)$ yields
\begin{equation}
E(z,f) = \tilde{E}_\text{LIN}(z,f) + E_\text{NLI}(z,f),
\end{equation}
where
\begin{equation}
\tilde{E}_\text{LIN}(z,f) = e^{\tilde{\Gamma}(z)}E_\text{in}(f)
\label{eq:elintilde}
\end{equation}
and
\begin{equation}
E_\text{NLI}(z,f) = e^{\tilde{\Gamma}(z)}\int_0^z dz' e^{-\tilde{\Gamma}(z')} Q_{\text{NLI},\tilde{A}_i}(z',f).
\label{eq:enli}
\end{equation}

Given the definition of $\hat{\tilde{\Gamma}}(z)$, the solution for $M(z,f)$ is given by Eq.~\eqref{eq:mdef}, where
\begin{align}
\tilde{\Gamma}(z) = \frac{1}{2}\left[(1-j\alpha_H)g_0 \left(1-\epsilon P_\text{in}\frac{z_\text{eff}(z)}{z}\right)-\alpha\right]z
\label{eq:tildegammadef}
\end{align}
and
\begin{equation}
z_\text{eff}(z) \Let \int_0^z dz' e^{(g_0 -\alpha)z'} = \frac{e^{(g_0 - \alpha)z}-1}{g_0 - \alpha}.
\label{eq:zeff}
\end{equation}
Note that the term containing $-\epsilon P_\text{in} z_\text{eff}(z)$ in $\tilde{\Gamma}$ is analogous to the term $-j2\gamma P_\text{in}z_\text{eff}(z)$ that arises in the perturbative treatment of the NLSE~\cite[Eq.~(47)]{Poggiolini2014detailed}. The latter, when exponentiated, simply gives rise to a distance-dependent phase rotation, which does not affect the noise power spectral density. Here, to the contrary, the term is relevant, because it leads to a reduction of the material gain coefficient from its small-signal value $g_0$, and therefore accounts for nonlinear gain compression.

However, the gain compression described by~\eqref{eq:tildegammadef} is incomplete. As shown in Appendix~\ref{app:gaincompression}, the compressed material gain coefficient is~\eqref{eq:gcompression} $g(z)=g_0/(1+\epsilon P(z))$, while according to~\eqref{eq:tildegammadef}, $g(z)=g_0(1-\epsilon P_\text{in}z_\text{eff}/z)$. These two expressions only agree in the limit $P(z)\ll P_\text{sat}$ and for small $z$, where $P(z)\approx P_\text{in}$ and $z_\text{eff}/z\approx 1$, so that in the former case $g(z)= g_0/(1+\epsilon P(z))\approx g_0(1-\epsilon P_\text{in})$.

The reason for this discrepancy is that the ratio $P(z)/P_\text{sat}$ is not a small parameter, and therefore gain compression cannot be described in linear order of the perturbation theory.
$Q_\text{NLI}$~\eqref{eq:qnlidef} was truncated at first order in $\epsilon$, while deterministic terms arising from the treatment of $Q_{\text{NLI},X_i}$ occur at all orders of the perturbation theory. To take them into account requires the summation of an infinite partial perturbation series, which is performed below.

In order to see how the infinite series arises and can be summed, consider the second-order term of the expansion~\eqref{eq:secondorder} of $\Delta g E$ in $\epsilon$.
Inserting the signal model~\eqref{eq:signalmodel} into this term and defining $f_m=m\Delta f$ etc. yields
\begin{align}
&\epsilon^2 g_0\!\int\! df_1\ldots df_4 E(f_1) E^*(f_1-f_2) E(f_3) E^*(f_3-f_4+f_2)\notag\\
&H_c(f_4-f_2) H_c(f_4) E(z,f-f_4)\notag\\
=&\epsilon^2 g_0(\Delta f)^\frac{5}{2}\!\!\!\!\!\!\sum_{m,n,k,r,s}\!\!\!\! \xi_m \xi_n^* \xi_k \xi_r^* \xi_s\!\! \int\! df_1\!\ldots\! df_4 \sqrt{G(f_1)G(f_1-f_2)}\notag\\
&\sqrt{G(f_3)G(f_3-f_4+f_2)G(f-f_4)} H_c(f_4-f_2)H_c(f_4)\notag\\
&e^{\Gamma(z)} e^{\Gamma^*(z)} e^{\Gamma(z)} e^{\Gamma^*(z)} e^{\Gamma(z)}\delta(f_1-f_m)\delta(f_1-f_2-f_n)\notag\\
&\delta(f_3-f_k)\delta(f_3+f_2-f_4-f_r)\delta(f-f_4-f_s).
\end{align}
Performing the integrals and accounting for the delta distributions imposes the following conditions: $f_1=f_m$, $f_2=f_m-f_n$, $f_3=f_k$, $f_4=f_3+f_2-f_r=f_k+f_m-f_n-f_r$. This yields $f-f_4-f_s = f-f_k-f_m+f_n+f_r-f_s$ (energy conservation) and $f_4-f_2 = f_k - f_r$. One obtains
\begin{align}
& \epsilon^2 g_0(\Delta f)^\frac{5}{2}\!\!\!\!\!\sum_{m,n,k,r,s}\!\!\!\!\! \xi_m \xi_n^* \xi_k \xi_r^* \xi_s \sqrt{G(f_m)G(f_n)G(f_k)G(f_r)G(f_s)}\notag\\
&\qquad H_c(f_k-f_r)H_c(f_k+f_m-f_n-f_r) \notag\\
&\qquad e^{4\text{Re}\Gamma(z)} e^{\Gamma(z)}\delta(f-f_m+f_n-f_k+f_r-f_s).
\end{align}
The dominant contributions to these sums stem from the frequency combinations for which the argument of the $H_c$ terms is zero. These are obtained for $k=r$ and $m=n$. In this case, the previous expression reduces to
\begin{align}
& \epsilon^2 g_0\,(\Delta f)^\frac{5}{2}\sum_{m,k,s} |\xi_m|^2 |\xi_k|^2 \xi_s G(f_m)G(f_k)\sqrt{G(f_s)}\notag\\
&\times e^{4\text{Re}\Gamma(z)} e^{\Gamma(z)}\delta(f-f_s) = \epsilon^2 g_0\left(P_\text{in}e^{2\text{Re}\Gamma(z)}\right)^2 E_\text{LIN}(z,f),
\end{align}
where~\eqref{eq:ptildee} was used as well as its equivalence to~\eqref{eq:ptx}.
These terms are precisely the deterministic contributions to $Q_\text{NLI}$ at second order in $\epsilon$. Alternatively, this result can be obtained by singling out those terms in the perturbation expansion~\eqref{eq:secondorder} for which the argument of the $H_c$ terms is zero.
This is the case for $f_2=0$ for the first-order contribution and for $f_2=f_4$ and $f_4=0$ for the second order. Singling out these contributions by formally replacing $H_c(f_2)=\delta(f_2)$, $H_c(f_4)=\delta(f_4)$, $H_c(f_4-f_2)=\delta(f_4-f_2)$ and performing the integrals yields
\begin{align}
&-\epsilon g_0 \int df_1 E(z,f_1) E^*(z,f_1) E(z,f)\notag\\
&+ \epsilon^2 g_0\! \int\!\! df_1 E(z,f_1) E^*(z,f_1)\! \int\!\! df_3 E(z,f_3) E^*(z,f_3) E(z,f).
\end{align}
When the field is replaced by the linear field, $E_\text{LIN}(z,f) = e^{\Gamma(z)} E(0,f)$, this becomes
\begin{align}
\left[-\epsilon g_0 P_\text{in}e^{2\text{Re}\Gamma(z)} + \epsilon^2 g_0 \left(P_\text{in}e^{2\text{Re}\Gamma(z)}\right)^2\right] E_\text{LIN}(z,f).
\end{align}
The expansion can be performed up to all orders, giving rise to the series
\begin{align}
g_0\left[\sum_{n=1}^\infty (-1)^n \epsilon^n \left(P_\text{in}e^{2\text{Re}\Gamma(z)}\right)^n\right] E_\text{LIN}(z,f).
\label{eq:infiniteorder}
\end{align}
The series includes the deterministic contributions proportional to $E_\text{LIN}(z,f)$ up to all orders in $Q_{\text{NLI},X_i}(z,f)$. Inserting this into the propagation equation~\eqref{eq:dedzfull} for the field yields
\begin{align}
&\frac{\partial}{\partial z}E(z,f) = -\frac{1}{2}\alpha E(z,f) + \frac{1}{2}(1-j\alpha_H)g_0\times\notag\\
&\left[1 - \epsilon P_\text{in}e^{2\text{Re}\Gamma(z)} + \left(\epsilon P_\text{in}e^{2\text{Re}\Gamma(z)}\right)^2 \mp\ldots\right] E(z,f)\notag\\
&+Q_{\text{NLI},\tilde{A}_i}(z,f),
\end{align}
which has the form
\begin{align}
\frac{\partial}{\partial z}E(z,f) = \hat{\Gamma}' E(z,f) + Q_{\text{NLI},\tilde{A}_i}(z,f).
\end{align}
The computation of the NLI field~\eqref{eq:enli} requires the evaluation of $\Gamma'(z)=\int_0^zdz'\hat{\Gamma}'(z')$, which yields
\begin{align}
\Gamma'(z) =& -\frac{1}{2}\alpha z + \frac{1}{2}(1-j\alpha_H)g_0\notag\\
& \int_0^z dz'\left[1 - \epsilon P_\text{in}e^{(g_0-\alpha)z} + \left(\epsilon P_\text{in}e^{(g_0-\alpha)z}\right)^2 \mp\ldots\right]\times\notag\\
=& -\frac{1}{2}\alpha z+ \frac{1}{2}(1-j\alpha_H)g_0 z + \frac{1}{2}(1-j\alpha_H)g_0\frac{1}{g_0-\alpha}\notag\\
&\left[- \epsilon P_\text{in}\frac{e^{(g_0-\alpha)z}-1}{1} + \left(\epsilon P_\text{in}\right)^2 \frac{e^{2(g_0-\alpha)z}-1}{2}\mp\ldots\right].
\end{align}
One recognizes the Taylor expansion of the natural logarithm in angular brackets, namely
\begin{align}
\log(1+x) = \sum_{n=1}^\infty (-1)^{n+1}\frac{x^n}{n},
\end{align}
(for $|x|<1$), so that
\begin{align}
\Gamma'(z)=&-\frac{1}{2}\alpha z+ \frac{1}{2}(1-j\alpha_H)g_0 z +\frac{1}{2}(1-j\alpha_H)\frac{g_0}{g_0-\alpha}\notag\\
&\times\left[-\log\left(1+\epsilon P_\text{in}e^{(g_0-\alpha)z}\right) + \log\left(1 + \epsilon P_\text{in}\right)\right].
\end{align}
Noting that typically $\epsilon P_\text{in} = P_\text{in}/P_\text{sat}\ll 1$ and further neglecting the waveguide attenuation $\alpha$ relative to $g_0$, one obtains
\begin{align}
\Gamma'(z) =&\frac{1}{2}(1-j\alpha_H)g_0 z 
- \frac{1}{2}(1-j\alpha_H)\log\left(1+\epsilon P_\text{in}e^{g_0 z}\right).
\label{eq:gammapfinal}
\end{align}
In order to compute the nonlinear interference field~\eqref{eq:enli}, substitute~\eqref{eq:gammapfinal} and evaluate $Q_{\text{NLI},\tilde{A}_i}(z',f)$ at the linear field $E'_\text{LIN}(z,f)=E(0,f)e^{\Gamma'(z)}$:
\begin{align}
&Q_{\text{NLI},\tilde{A}_i}[E'_\text{LIN}(z,f)]\! =\!-\frac{1}{2}(1\!-\!j\alpha_H)\epsilon g_0 (\Delta f)^{\frac{3}{2}}\sum_i \delta(f-i\Delta f)\notag\\
&\sum_{m,n,k\in \tilde{A}_i} \xi_m \xi_n^* \xi_k\sqrt{G_\text{in}(m\Delta f)G_\text{in}(n\Delta f)G_\text{in}(k\Delta f)}\notag\\
&\qquad\times H_c([m-n]\Delta f)e^{\Gamma'(z)} e^{\Gamma'^*(z)}e^{\Gamma'(z)}.
\end{align}
According to~\eqref{eq:enli} one has
\begin{align}
&E_\text{NLI}(z,f) = e^{\tilde{\Gamma}(z)} \int_0^z dz' e^{-\tilde{\Gamma}(z')}
Q_{\text{NLI},\tilde{A}_i}[E'_\text{LIN}(z,f)]\notag\\
&=-\frac{1}{2}(1-j\alpha_H)\epsilon g_0 (\Delta f)^{\frac{3}{2}}
\sum_{i=-\infty}^\infty \delta(f-i\Delta f)\!\!\!\!\!\!\sum_{m,n,k\in \tilde{A}_i} \xi_m \xi_n^* \xi_k\notag\\
&\quad\sqrt{G(m\Delta f)G(n\Delta f)G(k\Delta f)}H_c([m-n]\Delta f) \times\notag\\
&\quad e^{\Gamma'(z)} \int_0^z dz' e^{2\text{Re}\Gamma'(z')}
\label{eq:enli2}
\end{align}
The NLI field is of the form 
\begin{equation}
E_\text{NLI}(z,f) = \sum_i\mu_i\delta(f-i\Delta f)
\end{equation}
and its average power spectral density is
\begin{align}
G_{\text{NLI}} &= \langle |E_\text{NLI}(z,f)|^2\rangle = \sum_{ik}\langle \mu_i\mu_k^*\rangle\delta(f-i\Delta f)\delta(f-k\Delta f)\notag\\
&= \sum_{ik}\langle \mu_i\mu_k^*\rangle\delta(f-i\Delta f)\delta_{ik} = \sum_{i}\langle|\mu_i|^2\rangle\delta(f-i\Delta f).
\label{eq:gnli0}
\end{align}
Inserting expression~\eqref{eq:enli2} for the NLI field, one obtains
\begin{align}
G_{\text{NLI}} =& \frac{1}{4} (1+\alpha_H^2)\epsilon^2 g_0^2 (\Delta f)^3\times\notag\\
& \sum_{m,n,k\in \tilde{A}_i}\sum_{m',n',k'\in \tilde{A}_i} \langle \xi_m \xi_n^* \xi_k \xi_{m'}^* \xi_{n'} \xi_{k'}^*\rangle\times\notag\\
&\sqrt{G_\text{in}(m\Delta f)G_\text{in}(n\Delta f)G_\text{in}(k\Delta f)}H_c([m-n]\Delta f)\notag\\
&\sqrt{G_\text{in}(m'\Delta f)G_\text{in}(n'\Delta f)G_\text{in}(k'\Delta f)}H_c^*([m'-n']\Delta f)\notag\\
&\times e^{2\text{Re}\Gamma'(z)}
\left|\int_0^z dz' e^{2\text{Re}\Gamma'(z')}\right|^2.
\label{eq:gnli1}
\end{align}
It turns out that most of the products of random variables $\xi$ average to zero. Among all possible index combinations, the only averages that are relevant in the limit $\Delta f\to 0$ are those with~\cite{Poggiolini2014detailed}
\begin{align*}
m' &= m, n'=n, k'=k, m\neq n, n\neq k, m\neq k\\
k' &= m, n'=n, m'=k, m\neq n, n\neq k, m\neq k
\end{align*}
in which case
\begin{align}
\langle \xi_m \xi_n^* \xi_k \xi_{m'}^* \xi_{n'} \xi_{k'}^*\rangle = \langle |\xi_m|^2 \rangle \langle |\xi_n|^2 \rangle \langle |\xi_k|^2 \rangle = 1.
\end{align}
Using $n=m+k-i$ [see~\eqref{eq:aidef}], this results in
\begin{align}
G_\text{NLI}(z,f) =& 
\frac{1}{4}\left(1+\alpha_H^2\right) 
\epsilon^2 g_0^2 (\Delta f)^3\sum_{i=-\infty}^\infty \delta(f-i\Delta f)\notag\\
&\sum_{m}\sum_k G_\text{in}(f_m)G_\text{in}(f_m-f_i+f_k)G_\text{in}(f_k)\times \notag\\
& \times\left(|H_c(f_i-f_k)|^2 + H_c(f_i-f_k) H_c^*(f_i-f_m)\right)\notag\\
& e^{2\text{Re}\Gamma'(z)}
\left|\int_0^z dz' e^{2\text{Re}\Gamma'(z')}\right|^2.
\label{eq:gnli2}
\end{align}
Now define the normalized spectral density 
\begin{align}
g_\text{in}(f)\Let G_\text{in}(f)/P_\text{in},
\end{align}
in terms of which the nonlinear noise power spectral density becomes
\begin{align}
G_\text{NLI}(z,f) =& 
\frac{1}{4}\left(1+\alpha_H^2\right) 
(\Delta f)^3\sum_{i=-\infty}^\infty \delta(f-i\Delta f)\notag\\
&\sum_{m}\sum_k  g_\text{in}(f_m)g_\text{in}(f_m-f_i+f_k)g_\text{in}(f_k)\times \notag\\
&\left(|H_c(f_i-f_k)|^2 + H_c(f_i-f_k) H_c^*(f_i-f_m)\right)\notag\\
&P_\text{in} e^{2\text{Re}\Gamma'(z)}
\epsilon^2 \left|g_0 P_\text{in}\int_0^z dz' e^{2\text{Re}\Gamma'(z')}\right|^2.
\label{eq:gnli3}
\end{align}
Replacing $\Gamma'(z)$ by~\eqref{eq:gammapfinal} above gives rise to a factor
\begin{align}
P_\text{in} e^{2\text{Re}\Gamma'(z)} =  \frac{P_\text{in}e^{g_0 z}}{1+\epsilon P_\text{in}e^{g_0 z}}.
\label{eq:pinren}
\end{align}
The factor $1/(1+\epsilon P_\text{in}e^{g_0 z})$ is reminiscent of gain compression as in~\eqref{eq:gcompression}. Here, however, the growth of the input power is governed by the uncompressed gain coefficient $g_0$. Indeed, the growth equation
\begin{equation}
\frac{d P}{d z} = g_0 P(z),
\label{eq:dpdzg0p}
\end{equation}
has the solution
\begin{equation}
P(z)=P_\text{in}e^{g_0 z}.
\label{eq:p0z}
\end{equation}
This expression is analogous to~\cite[Eq. (44)]{Poggiolini2014detailed}, namely $P(z)=P_\text{in}e^{-\alpha z}$. While $\alpha$ in the NLSE remains constant, the gain coefficient in the Agrawal model is compressed according to~\eqref{eq:gcompression} when the signal is amplified and power increases. The spurious result~\eqref{eq:gnli3} is a consequence of the fact that~\eqref{eq:gammapfinal} has been derived based on the linear field subject to~\eqref{eq:elin1}, which  evolves according to the uncompressed gain.
To remedy this shortcoming and to properly account for gain compression in the evolution of $P(z)$, $g_0$ in~\eqref{eq:dpdzg0p} must be replaced by $g(z)$ according to~\eqref{eq:gcompression}, i.e.:
\begin{equation}
\frac{d P}{dz} = g(z) P(z).
\label{eq:dpdzgp}
\end{equation}
As shown in Appendix~\ref{app:gaincompression}, this equation indeed gives rise to the familiar gain compression in the Agrawal model.
The above substitution amounts to replacing the evolution~\eqref{eq:p0z} in~\eqref{eq:gnli3} by the actual one governed by gain compression:
\begin{align}
P_\text{in}e^{g_0 z} \to P(z)
\end{align}
and correspondingly 
\begin{align}
P_\text{in} e^{2\text{Re}\Gamma'(z)} \to \frac{P(z)}{1+\epsilon P(z)},
\end{align}
where $P(z)$ evolves with distance $z$ according to~\eqref{eq:dpdzgp}, where $g(z)=g(P(z))=g_0/(1+\epsilon P(z))$ according to~\eqref{eq:gcompression}. With these substitutions, one can write
\begin{align}
g_0 P_\text{in}\int_0^z dz' e^{2\text{Re}\Gamma'(z')} &= \int_0^z dz' \frac{g_0}{1+\epsilon P(z')} P(z').
\end{align}
Using again that $g(z)=g_0/(1+\epsilon P(z))$ and~\eqref{eq:dpdzgp}, one obtains 
\begin{align}
g_0 P_\text{in}\int_0^z dz' e^{2\text{Re}\Gamma'(z')} &= \int_0^z dz' g P(z') = \int_0^z dz' \frac{\partial P(z')}{\partial z'}\notag\\
&= P(z) - P_\text{in} = P(z)\left(1-\frac{1}{G(z)}\right),
\end{align}
where $G(z)\Let P(z)/P_\text{in}$. Defining $P_\text{out}\Let P(L)$ and $G\Let G(L)$, the noise power spectral density at the amplifier output at $z=L$ becomes
\begin{align}
G_\text{NLI}(L,f) =& 
\frac{1}{4}\left(1+\alpha_H^2\right) 
(\Delta f)^3\sum_{i=-\infty}^\infty \delta(f-i\Delta f)\notag\\
&\sum_{m}\sum_k  g_\text{in}(f_m)g_\text{in}(f_m-f_i+f_k)g_\text{in}(f_k)\times \notag\\
&\left(|H_c(f_i-f_k)|^2 + H_c(f_i-f_k) H_c^*(f_i-f_m)\right)\notag\\
&\frac{P_\text{out}}{1+\epsilon P_\text{out}}
\epsilon^2 P_\text{out}^2\left(1-\frac{1}{G}\right)^2.
\end{align}
Performing the continuum limit $\Delta f\to 0$ and replacing $\epsilon\to 1/P_\text{sat}$, one finally arrives at
\begin{align}
G_\text{NLI}(f) =& 
\frac{1}{4}\left(1+\alpha_H^2\right) \frac{P_\text{out}}{1+\frac{P_\text{out}}{P_\text{sat}}}\left(\frac{P_\text{out}}{P_\text{sat}}\right)^2\left(1-\frac{1}{G}\right)^2\times \notag\\
&\times\!\int_{-\infty}^{+\infty}\!\!\! df_1 \int_{-\infty}^{+\infty}\!\!\! df_2\, g_\text{in}(f_1)g_\text{in}(f_2)g_\text{in}(f_1+f_2-f)\notag\\
&\times\left(|H_c(f-f_2)|^2 + H_c(f-f_2) H_c^*(f-f_1)\right).
\label{eq:app:gnlifinal}
\end{align}
Here $G\Let P_\text{out}/P_\text{in}$ is the compressed gain, which is the solution to~\eqref{eq:gstatic}.

\section{Derivation of closed-form expression}
\label{sec:app:closedformideal}

In the following, a closed-form expression for the nonlinear NSR of a WDM signal with an idealized rectangular input power spectral density (ideal Nyquist-WDM) is derived.
Consider a rectangular power spectral density of bandwidth $B$.
The double integral in the noise power spectral density~\eqref{eq:app:gnlifinal} is evaluated at a frequency $f=f_0$ that is assumed to be not too close to the band edge, i.e., at a distance significantly larger than the cutoff frequency $f_c$~\eqref{eq:fc}. 
Without loss of generality, one may choose $f_0=0$ to be in the center of the band. 
The input power spectral density equals $g_\text{in}(f)=1/B$ for $f\in[f_0-B/2,f_0+B/2]$ and $g_\text{in}(f)\equiv 0$ elsewhere.
The integral is to be performed over a region with the shape of an elongated hexagon where the product $g_\text{in}(f_1)g_\text{in}(f_2)g_\text{in}(f_1+f_2)$ is non-zero (shaded region in Figs.~\ref{fig:integrala} and~\ref{fig:integralb} in the main text). The functions $|H_c(f-f_2)|^2$, $H_c^*(f-f_1)$ and $H_c(f-f_2)$ decay rapidly with frequency since the ratio of $f_c$ to the bandwidth is typically small (equivalently, $B\tau_c$ is reasonably large). For the same reason, edge effects are confined to within a narrow strip of width of order $f_c$ around the band edges. Due to this decay, the integration region can be extended to a square to good approximation, for both terms in the last line of~\eqref{eq:app:gnlifinal}, as illustrated in Figs.~\ref{fig:integrala} and~\ref{fig:integralb}. This extension allows one to obtain the integral in closed form. 
With these simplifying assumptions, the double integral in~\eqref{eq:app:gnlifinal} becomes
\begin{align}
I=& \frac{1}{B^3}\int_{f_0-B/2}^{f_0+B/2} df_2 |H_c(f-f_2)|^2 \int_{f_0-B/2}^{f_0+B/2} df_1\notag\\
& + \frac{1}{B^3}\int_{f_0-B/2}^{f_0+B/2} df_2 H_c(f-f_2) \int_{f_0-B/2}^{f_0+B/2} df_1 H_c^*(f-f_1).
\label{eq:app:I}
\end{align}

Using the identities
\begin{align}
\int_{f_0-B/2}^{f_0+B/2} \frac{1}{1+4\pi^2 \tau_c^2 (f_0-f_2)^2} df_2 &= \frac{\arctan(\pi B\tau_c)}{\pi \tau_c},\\
\int_{f_0-B/2}^{f_0+B/2} \frac{1}{1 \pm 2\pi i \tau_c (f_0-f_2)} df_2 &= \frac{\arctan(\pi B\tau_c)}{\pi \tau_c},
\end{align}
one obtains
\begin{align}
I&=\frac{1}{B}\frac{\arctan(\pi B\tau_c)}{\pi B\tau_c} + \frac{1}{B} \left(\frac{\arctan(\pi B\tau_c)}{\pi B\tau_c}\right)^2.
\end{align}
For a WDM channel of bandwidth $B_\text{ch}$, the nonlinear NSR is given by $G_\text{NLI}(f_0) B_\text{ch} / P_\text{ch}$, where $N_\text{ch}B_\text{ch} = B$ and $N_\text{ch}P_\text{ch} = P_\text{out}$. This leads to 
\begin{align}
\text{NSR}_\text{NL}\! =&\! \frac{1}{4}\!\left(1+\alpha_H^2\right)\! \frac{1}{1+\frac{P_\text{out}}{P_\text{sat}}}\left(\frac{P_\text{out}}{P_\text{sat}}\right)^2\!\left(1-\frac{1}{G}\right)^2\!\notag\\
&\times\left[\frac{\arctan(\pi B\tau_c)}{\pi B\tau_c} + \left(\frac{\arctan(\pi B\tau_c)}{\pi B\tau_c}\right)^2\right].
\label{eq:app:nsrwdm1}
\end{align}
Using approximation~\eqref{eq:approxatan} leads to~\eqref{eq:nsrwdm1}.

\section{Nonlinear NSR for spectrally shaped WDM signals}
\label{sec:app:closedformshaped}

It is instructive to consider the effect of a non-rectangular spectral shape of the input PSD $G_\text{in}(f)$. We assume that the spectral width of the SOA filter is sufficiently narrow and smaller than the spectral gap between channels. Such approximation is more accurate for larger $\tau_c$ and higher baud rates. 
The integral in~\eqref{eq:app:gnlifinal} can be rewritten in the form
\begin{align}
I'=&\int_{-\infty}^{+\infty}\!\!\! df_1 \int_{-\tilde{B}/2}^{+\tilde{B}/2}\!\!\! d\Delta f g_\text{in}(f_1)g_\text{in}(f-\Delta f)g_\text{in}(f_1-\Delta f)\notag\\
&\times\left(|H_c(\Delta f)|^2 + H_c(\Delta f) H_c^*(f-f_1)\right).
\end{align}
Here $\tilde{B}$ is a fictitious bandwidth which is chosen large enough such that the integral essentially covers the whole SOA bandwidth (i.e., it should exceed the cutoff frequency $f_c$) and small enough such that $g_\text{in}(\Delta f)$ can be considered constant over $\tilde{B}$.
These contradictory requirements can be better fulfilled the larger $\tau_c$, which permits a smaller $\tilde{B}$. For sufficiently small $\tilde{B}$, $\Delta f$ can be neglected in the arguments of $g_\text{in}$ and we have
\begin{align}
I'\approx & g_\text{in}(f)\int_{-\infty}^{+\infty}\!\!\! df_1 g_\text{in}^2(f_1) \int_{-\tilde{B}/2}^{+\tilde{B}/2}\!\!\! d\Delta f \notag\\
&\times\left(|H_c(\Delta f)|^2 + H_c(\Delta f) H_c^*(f-f_1)\right).
\end{align}
The spectral shape of the noise hence approximately follows that of the WDM signal, similarly to the case of fiber nonlinear noise~\cite{Poggiolini2014}.
The second term containing $H_c^*(f-f_1)$ is generally smaller than the first and significant only in case of self-channel interference, and only when $f_1$ is close to $f$. In this term we make the additional approximation that the argument of $g_\text{in}^2(f_1)$ can be approximated by $f$. This yields
\begin{align}
I'\approx & g_\text{in}(f)\int_{-\infty}^{+\infty}\!\!\! df_1 g_\text{in}^2(f_1) \int_{-\tilde{B}/2}^{+\tilde{B}/2}\!\!\! d\Delta f |H_c(\Delta f)|^2\notag\\
& + g_\text{in}(f)^3\int_{-\infty}^{+\infty}\!\!\! df_1 H_c^*(f-f_1) \int_{-\tilde{B}/2}^{+\tilde{B}/2}\!\!\! d\Delta f  H_c(\Delta f).
\end{align}
If we assume the power spectrum of the WDM signals to be raised-cosine shaped with roll-off factor $\beta$, the calculation of the NSR involves the integrals
\begin{equation}
\rho_k\Let\frac{1}{R_s}\int_{-\infty}^{+\infty}\!\!\! \text{rc}^k(f)df  =1-\beta\left(1-\frac{2\sqrt{\pi}\sec(k\pi)}{\Gamma(\frac{1}{2}-k)\Gamma(1+k)}\right).
\label{eq:app:rck}
\end{equation}
It follows that $\rho_1=1$ and $\int_{-\infty}^{+\infty} \text{rc}(f)df = R_s$. For a signal with $N_\text{ch}$ channels at center frequencies $f_{c,n}$ the normalized input power spectral density reads
\begin{equation}
g_\text{in}(f) = \frac{1}{N_\text{ch}}\frac{1}{R_s}\sum_{n=1}^{N_\text{ch}} \text{rc}(f-f_{c,n}).
\label{eq:app:wdmpsd}
\end{equation}
Evaluating $I'$ similarly to~\eqref{eq:app:I} and $\text{NSR}_\text{NL}$ as $(1/P_\text{ch})\int_{-B_\text{ch}/2}^{B_\text{ch}/2} df G_\text{NLI}(f)$, we finally obtain
\begin{align}
\text{NSR}_\text{NL}\! =&\! \frac{1}{4}\!\left(1+\alpha_H^2\right)\! \frac{1}{1+\frac{P_\text{out}}{P_\text{sat}}}\left(\frac{P_\text{out}}{P_\text{sat}}\right)^2\!\left(1-\frac{1}{G}\right)^2\notag\\
&\times
\left[\mu\frac{1}{2 N_\text{ch}R_s\tau_c}+\nu\left(\frac{1}{2 N_\text{ch}R_s\tau_c}\right)^2\right],
\label{eq:app:nsrrrc}
\end{align}
where $\mu = \rho_1\rho_2= 1-\beta/4$ and $\nu=\rho_3 = 1-3\beta/8$. 
Compared to the closed-form result~\eqref{eq:gnlifinal} for ideal Nyquist-WDM, the bandwidth is replaced by the occupied bandwidth $N_\text{ch}R_s$.
Note that this NSR expression does not account for receiver-side matched filtering. If filtering by $H_\text{RX}(f)$ is taken into account, the nonlinear NSR is given by
\begin{equation}
\text{NSR}_\text{NL} = \frac{1}{P_\text{ch}}\int_{-B_\text{ch}/2}^{B_\text{ch}/2}df G_\text{NLI}(f) |H_\text{RX}(f)|^2.
\end{equation}
For a root-raised cosine filter $H_\text{RX}(f)=\text{rrc}(f)$, the NSR is still given by~\eqref{eq:app:nsrrrc}, but with $\mu = \rho_2^2 = (1-\beta/4)^2$ and $\nu=\rho_4=1-29\beta/64$.

For an RRC, the filter changes from maximum to zero within a bandwidth of $2\beta R_s$, where $R_s$ is the sampling frequency. For $R_s=68$ GHz and roll-off $\beta=0.05$, $2\beta R_s = 6.8$ GHz. This value is not much larger than the typical cutoff frequency (1.6 GHz for $\tau_c=100$ ps), so that one may expect some degree of inaccuracy due to edge effects.

\begin{figure}[!t]
\centering
\includegraphics[width=0.93\columnwidth]{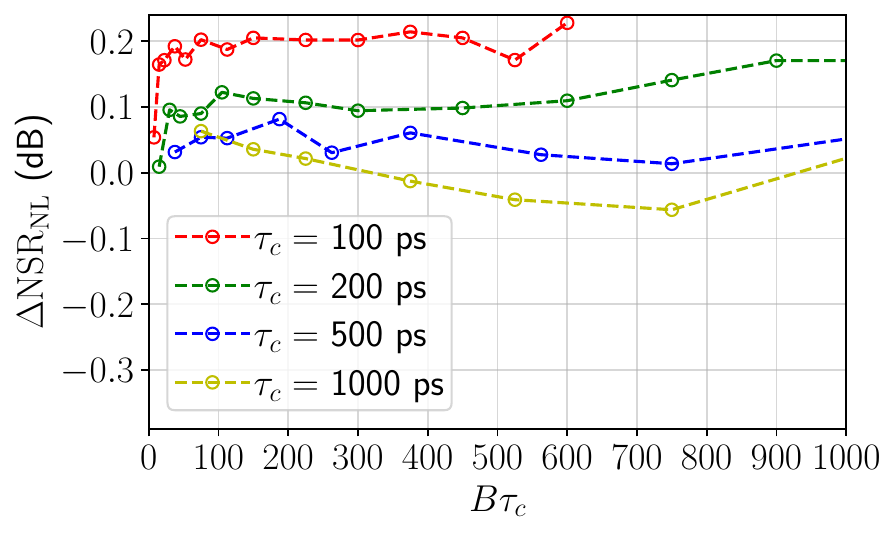}
\caption{NSR error of the closed-form GN model expression~\eqref{eq:app:nsrrrc} relative to numerical simulations of the Agrawal model where the number of channels is varied from $N_\text{ch}=1$ to $N_\text{ch}=80$ for fixed per channel power, such that $B=N_\text{ch}B_\text{ch}$ and total output power equals $P_\text{sat}=24$ dBm for the fully loaded band. Parameters are $B_\text{ch} = 75$ GHz, $R_s = 68$ GBaud, $G_0=10$ dB, $\alpha_H=5$ and raised-cosine shaped power spectral density is assumed with roll-off $\beta=0.05$.}
\label{fig:errnsrnlvsbt_tauc_rrc_apprx2}
\end{figure}

We assess the accuracy of the GN-model for a roll-off $\beta=0.05$. The noise power is computed from the waveform as before and receiver-side matched filtering is not taken into account. Fig.~\ref{fig:errnsrnlvsbt_tauc_rrc_apprx2} shows the error as a function of $B\tau_c$. It has a component that decays quickly with $B\tau_c$ and a systematic error that decays as $1/\tau_c$ independently of bandwidth. The latter is due to the edge effects. For sufficiently large values of $\tau_c$ (a few hundred ps), edge effects become  negligible. The formula is generally accurate, with the error not exceeding $\sim 0.2$ dB.

\section{Closed-form result for FWM efficiency}
\label{sec:app:fwmres}

The GN model formula~\eqref{eq:gnlifinal} can be interpreted as an integral over infinitesimal FWM products. It turns out that the FWM efficiency of continuous-wave (CW) light sources can be computed based on this formula, even though the assumption of Gaussian statistics is not satisfied.
In case of degenerate FWM, two CW sources are placed at frequencies $f_0$ and $f_0+\Delta f$ at SOA input, where $\Delta f$ is the frequency spacing. Ignoring the finite linewidth of actual lasers, the normalized input PSD for this case reads
\begin{equation}
g_\text{in}(f) = \frac{1}{2}\left[\delta(f-f_0) + \delta(f-f_0-\Delta f)\right].
\label{eq:gfwm}
\end{equation}
First-order four-wave mixing products are generated at the frequency of the lasers, as well as at $f=f_0-\Delta f$ and $f_0 +2\Delta f$. The FWM efficiency is defined as the ratio between the power of the first sideband to the power of the pump. Inserting the PSD into~\eqref{eq:gnlifinal} and taking the ratio of the sideband power at e.g. $f_0 +2\Delta f$ to the pump power, leads to the following closed-form result for the FWM efficiency 
\begin{align}
\text{FWM}(\Delta f) =& 
\frac{1}{32}\left(1+\alpha_H^2\right) \frac{1}{1+\frac{P_\text{out}}{P_\text{sat}}}\left(\frac{P_\text{out}}{P_\text{sat}}\right)^2\left(1-\frac{1}{G}\right)^2 \notag\\
&\times\frac{2}{1 + (\Delta f/f_c)^2}.
\label{eq:fwmeff}
\end{align}
Here it has been assumed that the power of the first sideband is negligible compared to the power of the pump (which hence remains undepleted). The result is based on first-order perturbation theory and therefore does not describe higher-order sidebands. They must hence be negligible (which is generally the case when the previous assumption is satisfied). Furthermore, the beating of two spectral tones leads to large amplitude oscillations between zero to twice the average power, for which a first-order perturbation theory may be inadequate. For these reasons, one may expect the applicability of the formula to be limited to low power.

\begin{figure}[!t]
\centering
\includegraphics[width=0.93\columnwidth]{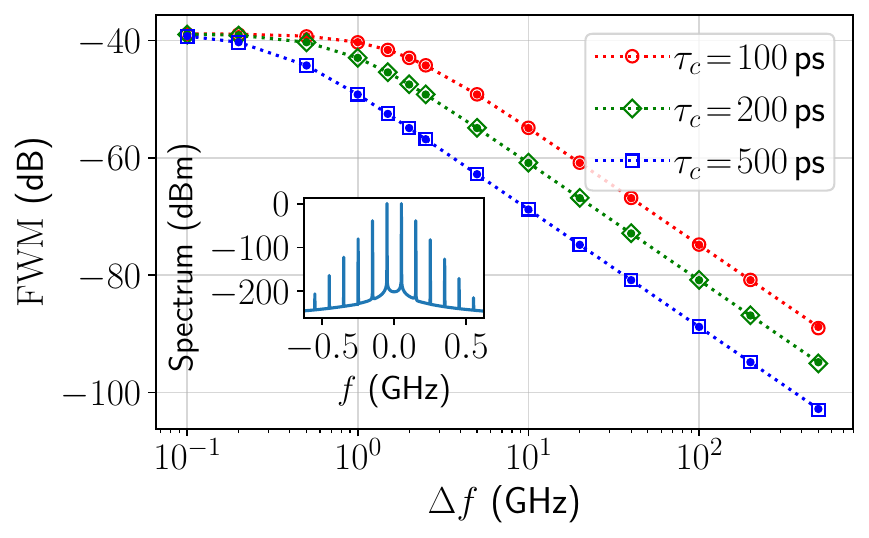}
\caption{Comparison of FWM mixing efficiency as predicted by formula~\eqref{eq:fwmeff} (lines with small points) compared to numerical simulation of the Agrawal model (open symbols) as a function of the frequency spacing $\Delta f$ for $P_\text{out} = P_\text{sat} - 20$ dB, $G_0 = 10$ dB and $\alpha_H=5$. The inset shows the FWM spectrum for $\Delta f = 0.1$ GHz and $\tau_c = 500$ ps.}
\label{fig:fwmvsdf_tauc_inset}
\end{figure}

\begin{figure}[!t]
\centering
\includegraphics[width=0.93\columnwidth]{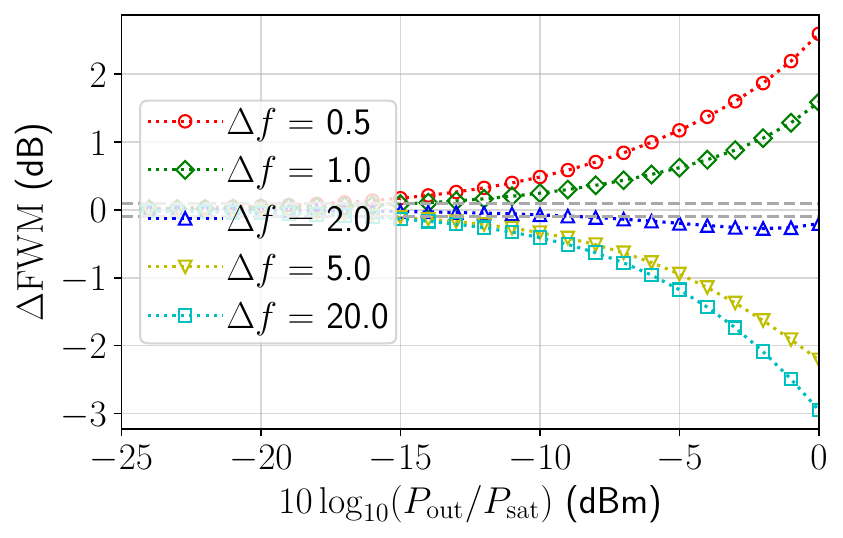}
\caption{Error of  FWM mixing efficiency as predicted by formula~\eqref{eq:fwmeff} relative to numerical simulations of the Agrawal model as a function of output power relative to $P_\text{sat}$ for different $\Delta f$ and for $G_0 = 10$ dB, $\alpha_H=5$, $\tau_c=100$ ps.}
\label{fig:fwmvsdf_fp_error_pout}
\end{figure}

It is nevertheless instructive to integrate the FWM efficiency over the frequency spacing. Using $\int_0^\infty d\Delta f \frac{2}{1+(\Delta f/f_c)^2} = \pi f_c = 1/(2\tau_c)$, one obtains
\begin{align}
&\int_0^\infty \text{FWM}(\Delta f) d\Delta f = \notag\\
&\qquad \frac{1}{32}\left(1+\alpha_H^2\right) \frac{1}{1+\frac{P_\text{out}}{P_\text{sat}}}\left(\frac{P_\text{out}}{P_\text{sat}}\right)^2\left(1-\frac{1}{G}\right)^2\frac{1}{2\tau_c}.
\end{align}
Interestingly, the integrated FWM efficiency is proportional to the nonlinear NSR of a WDM signal~\eqref{eq:nsrwdm3}: $\text{NSR}_\text{NL} = (8/B)\int_0^\infty \text{FWM}(\Delta f) d\Delta f$. At least at low power, where the formula is applicable, FWM experiments provide information about the NSR in a WDM system. 

The formula further explains the typical shape of experimental FWM curves (see, e.g., Ref.~\cite{Hafermann2024}).
Fig.~\ref{fig:fwmvsdf_tauc_inset} shows results for the FWM efficiency predicted by the formula~\eqref{eq:fwmeff} and Agrawal model simulations. 
Here $P_\text{out}$ was chosen 20 dB below $P_\text{sat}$. The power of the idler is thus $\sim 40$ dB below the power of the pump. The curves show a plateau at small frequency spacing $\Delta f\ll f_c$. The crossover to a power law decay is determined by the cutoff frequency $f_c$.
The formula agrees well with the simulation results, with an error of the order of 0.05 dB. It is independent of $\alpha$ and virtually independent of $\tau_c$. 
Fig.~\ref{fig:fwmvsdf_fp_error_pout} shows the error 
\begin{equation}
\Delta\text{FWM} = 10\log_{10}(\text{FWM}_\text{NL}^\text{GN} / \text{FWM}_\text{NL}^\text{SIM}) (\text{dB})
\label{eq:deltafwmdef}
\end{equation}
of the FWM efficiency for different frequency spacings, as a function of $P_\text{out}$. The error increases with $P_\text{out}$ as expected.
}

\bibliographystyle{IEEEtran}
\bibliography{references}

\end{document}